\documentclass[aps, prl, twocolumn, tightenlines, letterpaper, preprintnumbers, floatfix, longbibliography, nofootinbib,superscriptaddress,10pt]{revtex4-2}

\usepackage{amsmath,amssymb,mathrsfs,bm,graphicx,booktabs}
\usepackage[colorlinks=true, allcolors=blue]{hyperref}

\newcommand{\theff}{\theta_{\mathrm{eff}}}
\newcommand{\Ein}{E_{\mathrm{in}}}
\newcommand{\energy}{\mathscr{E}}
\newcommand{\Sconf}{S_{\mathrm{config}}}
\newcommand{\Ssec}{S_{\mathrm{sec}}}
\newcommand{\dbreak}{d_{\mathrm{break}}}

\newcommand{\DSasym}{\Delta S_{\mathrm{asym}}}

\usepackage{orcidlink}

\usepackage{xcolor}
\begin{document}
\setcounter{secnumdepth}{3}
\title{Electric Flux as a Criterion for String Breaking and Schwinger Pair Production}
\author{Sebastian Grieninger\,\orcidlink{0000-0002-9523-5819}}
\email{segrie@uw.edu}
\date{\today}
\affiliation{InQubator for Quantum Simulation (IQuS), Department of Physics, University of Washington, Seattle, WA 98195, USA}
\begin{figure}
\vskip -1.cm
\leftline{\includegraphics[width=0.15\textwidth]{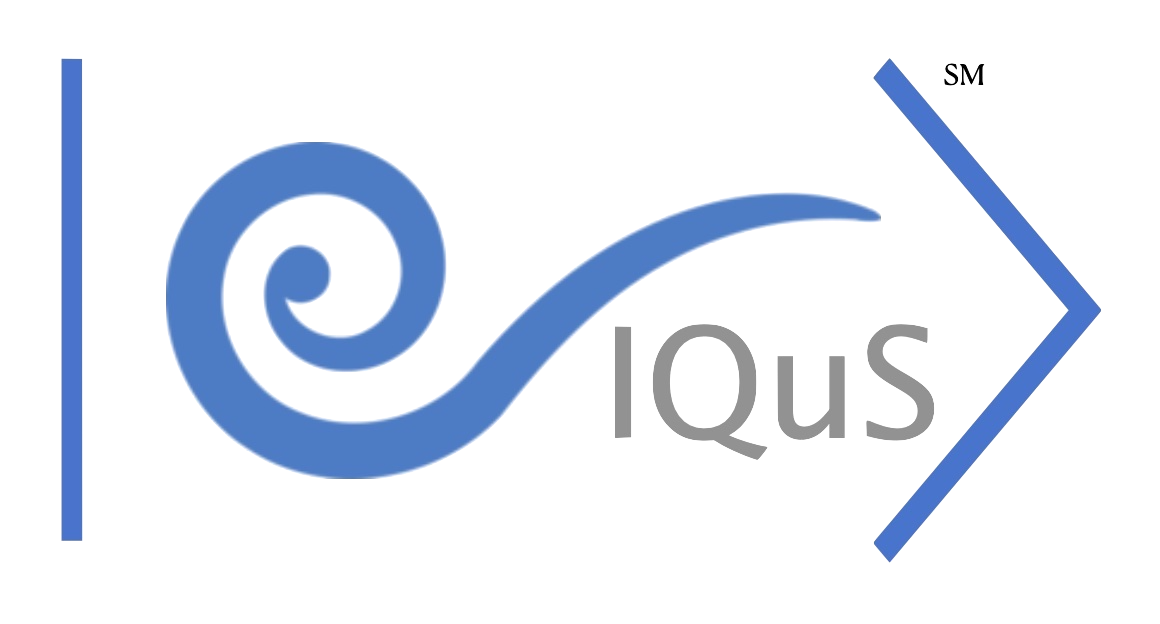}}
\end{figure}
\preprint{IQuS@UW-21-136, NT@UW-26-20}
\begin{abstract}
We define string breaking by the midpoint electric field falling to half a flux unit. 
In the massive Schwinger model, an energy balance between intact string and screened configuration predicts this separation using only vacuum observables and meson masses as input. We observe numerically that this half flux point corresponds to equal weights of the two configurations thus explaining the observed peak of the entanglement entropy as maximal uncertainty of the binary flux. The same binary uncertainty explains the entanglement peak at the vacuum transition.  Near breaking, screening organizes part of the spatial entanglement between the string interior and exterior into a nearly maximally entangled bit. Within the selected flux sectors, this approximate encoded Bell pair records whether the flux is screened and nearly separates from the remaining correlations. The vacuum transition exhibits the same structure. The string and the $\theta$-vacuum reveal a matching structure of the capacity of entanglement minimum at their transition point, linking their entanglement spectra. The transition rearranges degrees of freedom coherently as revealed by the entanglement asymmetry and Robustness of Magic of of disjoint subregions.  Finally, we construct the $\theta$-vacuum and string wavefunctions with high fidelity using a restricted Boltzmann machine ansatz.  
\end{abstract}
\maketitle
\section{Introduction}
\label{sec:introduction}

Quarks and gluons can only be observed through bound states like hadrons. One natural framework to study this experimentally are high-energy collisions where the produced quarks and gluons fragment into outgoing hadronic states. Understanding their formation  requires the nonperturbative physics of confinement.  

String breaking is the underlying process thus central to the description of fragmentation in high-energy collisions. The Lund model and its implementation in event generators  such as PYTHIA describe hadron formation through successive breakings of a color string by producing quark--antiquark pairs~\cite{Andersson:1983ia,Sjostrand:2019zhc}. 
A microscopic account of this process allows us to understand which (quantum) correlations are driving this process and have to be factored into phenomenological modeling. Studying string breaking with static external charges isolates the correlations from the motion of the sources.

In quantum chromodynamics (QCD), chromoelectric strings or flux tubes form between color charges at intermediate separations.  The energy stored between a quark and antiquark grows approximately linearly with their distance until the creation of a dynamical quark--antiquark pair becomes energetically favorable. Each static source can bind a dynamical partner into a static-light meson, leaving two color singlets in place of the flux tube. The ground state depends on the energies of both configurations and their mixing. Lattice QCD calculations have resolved this mixing in the low-energy spectrum~\cite{PhysRevD.59.031501,CP-PACS:1998hwq,SESAM1998209,Bali:2005fu,Pennanen:2000yk,PhysRevD.63.111501,Bulava:2019iut}.

The massive Schwinger model which is quantum electrodynamics in 1+1d provides a minimal realization of confinement and has recently gained a lot of attention due to its accessibility to Hamiltonian lattice simulations 
Hamiltonian lattice simulations using matrix product states have determined the profiles of classical observables such as charge density and the electric field of static strings together with their entanglement and quantum complexity structure%
~\cite{Buyens:2015tea, Grieninger:2025rdi,Grieninger:2026bdq}. Real-time calculations with moving external
charges examine the formation of fragmentation products and their
correlations~\cite{Florio:2023dke,Florio:2024aix,Florio:2025hoc}, while quantum
simulators have realized string breaking in one-dimensional lattice
gauge theories~\cite{Liu:2024lut,De:2024smi}. Further progress in higher dimensions, non-abelian gauge groups and quantum hardware has been achieved in~\cite{ Cochran:2024rwe,Gonzalez-Cuadra:2024xul, Borla:2025gfs,Cataldi:2025cyo,Xu:2025abo,DiMarcantonio:2025cmf,Ciavarella:2024fzw,Ciavarella:2024cyt,Halimeh:2021ufh,Zache:2021ggw,Barata:2025hgx,Artiaco:2025qqq,Verdel:2019chj,Verdel:2023mmp,Mallick:2024slg,Crippa:2024hso,Liu:2024lut,De:2024smi,Surace:2024bht,Ciavarella:2024lsp,Alexandrou:2025vaj,Verde:2026hcg,Zwolfer:2026yuc,Xu:2026ibi,Luo:2025qlg,John:2026gut,Batini:2024zst,Berges:2017hne}.

The mean electric field captures discharge by pair creation, but it averages over the different charge and flux configurations in the state leaving their correlations unresolved. Even ground states, subregions are mix after tracing out the environment due to their
entanglement with the exterior.
Kharzeev's maximal entanglement limit (MEL) connects the emergence
of a probabilistic description to phase information inaccessible
to the measured subsystem~\cite{Kharzeev:2026jkq}.
For static Schwinger strings, comparisons with thermal states
have explored this connection near breaking%
~\cite{Grieninger:2025rdi}. Further evidence for thermal behavior comes from~\cite{Berges:2017zws}. To identify the quantum correlations governing string breaking it is instructive to investigate the flux resolved entanglement spectrum. In particular, the majority of the observed peak in the entanglement entropy at breaking is rooted in the uncertainty about the electric flux, i.e. whether the string has broken or is intact. The rearrangement of quantum correlations within the fluxtube and the broken configuration can be investigated by flux projection. 

An exciting new way to think about quantum correlations comes from the quantum information science community. Non-classical correlations can be identified by deviations from so-called stabilizer states. A stabilizer state can have large entanglement while retaining an efficient classical description. Nonstabilizerness, or magic, quantifies a resource beyond this class in a specified
qubit encoding~\cite{Howard:2017maw,Leone:2021rzd,Falcao:2024msg,Robin:2026lqp,Robin:2025dgw,Keeble:2026rto,Robin:2025wip,Jha:2026eni,Magni:2026wfc,Sierant:2026lvg,Kadam:2026srm,Sierant:2026qji,Chu:2026yxm,Froland:2026aff,Iannotti:2026haw,Jasser:2026hrg,Ciavarella:2026tau,Bacciconi:2026qrs,Yao:2026rya,Amorosso:2026mdo,Ebner:2025pdm,Lee:2023urk,Viscardi:2026oos,Cao:2024nrx,Grieninger:2026txt}, motivating its study
alongside entanglement in Schwinger strings%
~\cite{Grieninger:2026bdq}.

\section{Electric flux and pair production}
\label{sec:model}
The Schwinger model is electrodynamics in one spatial dimension with massive Dirac fermions of mass $m$ and charge $g$. Pair production can be induced by coupling the Schwinger model to an external electric field and the resulting dynamical fermions screen the electric flux. We first ask how to identify the length at which a string breaks. The massless Schwinger model is exactly solvable which allows us to establish an analytical framework of the screening crossover. The description can be extended perturbatively to first order in the fermion mass. At small masses, however, the vacuum polarizes very easily and external charges are screened before a string forms. The solutions at finite mass are constructed numerically using a Hamiltonian lattice formulation. 
\subsection{Hamiltonian and static sources}

Bosonization expresses the
Schwinger model in terms of a scalar screening field $S$ and its conjugate
momentum $\Pi_S$. The free scalar $S$ is coupled to an external electric field $F_d(x)$ corresponding either to a $\theta$ angle, a pair of static charges, or both. The Hamiltonian density in bosonic form is:
\begin{align}
 \mathcal H={}&
 \frac{\Pi_S^2}{2\pi}
 +\frac{\pi}{2}(\partial_x S)^2
 +\frac{g^2}{2}(F_d-S)^2
 \nonumber\\
 &+m\Sigma_0\bigl[1-:\!\cos(2\pi S)\!:\!_\mu\bigr],
 \label{eq:bosonH}\\
 \mu={}&\frac{g}{\sqrt{\pi}},\qquad
 \Sigma_0=\frac{e^\gamma g}{2\pi^{3/2}} ,
 \label{eq:boson-normalization}
\end{align}
where $g$ is the electric charge, $\Sigma_0$ the free vacuum chiral condensate and $\gamma$ is Euler's constant.
The colon indicates normal ordering with respect to the free boson of
mass $\mu$~\cite{Coleman:1975pw}. 

The background electric field $\theta$ and the two static charges $+qg$ and $-qg$ at
$x=-d/2$ and $x=d/2$ are coupled to the Hamiltonian through the external source
\begin{equation}
 F_d(x)=\frac{\theta}{2\pi}+q f_d(x),
 \qquad
 f_d(x)=
 \begin{cases}
 1,& |x|<d/2,\\
 0,& |x|>d/2.
 \end{cases}
 \label{eq:source-profile}
\end{equation}
We write the physical electric field as $\mathcal E=gE=g(F_d-S)$. 
Thus $S$ is the screening of the external electric field due to the dynamical fermions. The electric term in Eq.~\eqref{eq:bosonH} favors $S=F_d$ 
and the mass terms is periodic under one unit of screening, i.e. $S\to S+1$.


The static potential describing the string is the difference of the ground-state energy with the two external charges $\energy_0(d)$  and the vacuum energy
$\energy_{\rm vac}$,
\begin{equation}
 V(d)=\energy_0(d)-\energy_{\rm vac}.
 \label{eq:pot}
\end{equation}
At large separation, each unit charge binds a dynamical screening
fermion forming a static-light meson. We can read off the total energy of these two mesons $M_H$ from the potential once they are well separated which is formally defined as
 $M_H=V(d\to\infty)$,
with the infinite-volume limit taken first. The vacuum meson mass $M_s$ refers to the lightest neutral excitation which is bound state of dynamical fermions. The mass $M_s$ sets the vacuum correlation length,
whereas $M_H$ includes the energy of the two screening clouds bound to
the static charges. 

\subsection{Lattice formulation}

For the numerical calculations we use a ring of circumference $L$,
with $N$ staggered sites and lattice spacing $a=L/N$. The separation of the external charges is denoted by $d$ corresponding to $d/a$ links. The Kogut--Susskind
Hamiltonian with staggered fermions is%
~\cite{Kogut:1974ag,Susskind:1976jm,Banks:1975gq}
\begin{align}
 H={}&-\frac{i}{2a}\sum_n
 \left(\chi_n^\dagger U_n\chi_{n+1}-{\rm h.c.}\right)
 +m_{\rm lat}\sum_n(-1)^n\chi_n^\dagger\chi_n
 \nonumber\\
 &+\frac{ag^2}{2}\sum_n
 \left(L_n+\frac{\theta}{2\pi}+q f_n(d)\right)^2 .
 \label{eq:H}
\end{align}
 Here $\chi_n$ is the staggered fermion
operator, $U_n$ raises the integer
electric flux $L_n$ on a link and $f_n$ is the external source
in Eq.~\eqref{eq:source-profile} and
$m_{\rm lat}=m-g^2a/8$ the improved lattice mass~\cite{Dempsey:2022nys}. The lattice electric field is
\begin{equation}
 \mathcal E_n=g\widehat E_n
 =g\left(L_n+\frac{\theta}{2\pi}+q f_n(d)\right).
 \label{eq:fielddef}
\end{equation}
Gauss's law
$L_n-L_{n-1}=\chi_n^\dagger\chi_n-[1-(-1)^n]/2$ allows to write the gauge field in terms of fermionic degrees of freedom up to one independent integer on the ring, $L_0$ which is conjugate to the gauge holonomy. Hopping across the periodic boundary changes this flux, allowing $L_0$ to change through dynamical charge transport.

We obtain the ground states and low-lying excitations with matrix product states. Exact diagonalization supplies the reference states for the neural quantum-state benchmarks in section~\ref{sec:neural-states}. We give further details on the numerical methods in Appendix~\ref{app:numerics}.

\subsection{Exact screening at zero fermion mass}
\label{sec:massless-screening}

At $m=0$ the cosine interaction in the Hamiltonian~\eqref{eq:H} vanishes. The external electric field then displaces the free massive boson, leaving its fluctuation frequencies unchanged. The exact
ground-state mean $S_0$ follows by minimizing the quadratic energy. Throughout this section, we set $\theta=0$.
On an infinite line the static equation of motion is
\begin{equation}
 (-\partial_x^2+\mu^2)S_0(x;d)=\mu^2 q f_d(x),
 \qquad S_0(\pm\infty;d)=0.
 \label{eq:massless-profile-equation}
\end{equation}
The associated Green's function is $G_0(x)=e^{-\mu|x|}/(2\mu)$. Using the explicit form of $f_d$ in~\eqref{eq:source-profile} yields the analytic expression for the screening field
\begin{equation}
 S_0(x;d)=q
 \begin{cases}
 1-e^{-\mu d/2}\cosh(\mu x),& |x|<d/2,\\
 \sinh(\mu d/2)e^{-\mu|x|},& |x|>d/2 .
 \end{cases}
 \label{eq:massless-profile}
\end{equation}
Between the external charges the electric field is consequently
\begin{equation}
 E(x)=q e^{-\mu d/2}\cosh(\mu x).
 \label{eq:elboson}
\end{equation}
The screening charge accumulates around each static charge and both clouds fall off exponentially. Their overlap leaves the residual electric field in the middle~\cite{Okuda:2022hsq}.

The potential includes both the electric and gradient energies of the
screening profile. Integrating the gradient term by parts and using the
static field equation gives
\begin{align}
 V(d,0)
 &=\frac{g^2}{2}\int dx\,F_d(x)\bigl[F_d(x)-S_0(x;d)\bigr]
 \nonumber\\
 &=\frac{q^2g^2}{2\mu}(1-e^{-\mu d}).
 \label{eq:massless-potential}
\end{align}
At small separations $d$ the potential has slope $q^2g^2/2$ which is the energy density of an unscreened electric field. At large separations it approaches the sum of the screening self-energies of the two static charges
\begin{equation}
 M_s(0)=\mu,\quad M_H(0)=\frac{g^2}{2\mu}
 =\frac{g\sqrt{\pi}}{2}, \quad (\text{for } q=1).
 \label{eq:massless-scales}
\end{equation}

\subsection{Flux branches in the vacuum and the string}
\label{sec:branches}
In the massless theory, the electric field is solely depleted by screening. At finite fermion mass the external charges can be connected by a flux tube which eventually breaks as they separate. We argue that string breaking is a transition between competing electric-flux configurations. We develop this picture by first examining the electric-flux configurations of the $\theta$-vacuum in the heavy-fermion limit.

The mean electric field follows from the vacuum energy density through
$\langle E\rangle=(2\pi/g^2)\,\partial_\theta\epsilon_{\rm vac}$.
In the heavy-fermion limit, only the electric term contributes to its
$\theta$ dependence, giving $E=\theta/(2\pi)-n$ for the
ground state at reference flux $L_0=-n$. Up to a common energy density,
the branch energies therefore approach
\begin{equation}
 \bar\epsilon_n(\theta)\longrightarrow
 \frac{g^2}{2}\left(\frac{\theta}{2\pi}-n\right)^2,
 \qquad m/g\to\infty .
 \label{eq:heavy-vacuum-branches}
\end{equation}
The energy difference of the $n=0$ and $n=1$ branches is
\begin{equation}
 \bar\epsilon_1(\theta)-\bar\epsilon_0(\theta)
 \longrightarrow \frac{g^2}{2}\left(1-\frac{\theta}{\pi}\right).
 \label{eq:heavy-vacuum-branch-difference}
\end{equation}
The energy ordering reverses at $\theta=\pi$, where the two competing branches carry flux $+1/2$ and $-1/2$. Increasing $\theta$ by $2\pi$ and shifting $n$ by one leaves the energy unchanged and the vacuum energy is periodic.

At finite mass, the mean electric field reverses continuously as $\theta$ passes through $\pi$ below $m_c/g\simeq0.3335$ and discontinuously across a first-order transition above $m_c$~\cite{Coleman:1976uz,Byrnes:2002gj}.

The string carries a similar branch structure restricted to a finite spatial
interval. The region between the external charges is effectively shifted by $\theta\to \theta+2\pi q$ while the outside remains at $\theta$. Separating the sources increases the energy stored between them and hence the energy gained by screening it. Once this gain outweighs the energy required to produce a dynamical pair, the two configurations reverse their energy ordering and the string breaks.

A useful comparison for observables are fractional $q$ which have finite asymptotic string tension for $m>0$ (assuming $\theta=0$)~\cite{Gross:1995bp,Adam:1996rd}.
The special case is $q=1/2$ above $m_c$ where the string does not break since a dynamical pair reverses the interior electric field without reducing its energy density.

\subsection{Where does the string break?}
\label{sec:breaking-markers}

We follow $\Ein(d)$, the vacuum-subtracted electric field at the midpoint between the external charges. $\Ein(d)$ plays the role of an order parameter, parametrizing  string breaking similarly to $\theta$ in the vacuum. For $q=1$, we define the breaking separation by
\begin{equation}
 \Ein(\dbreak)=\frac12 .
 \label{eq:breaking-criterion}
\end{equation}
The half-flux value lies halfway between the unscreened intact string and the broken fully screened configuration.

Note that the midpoint electric field is the first moment of a local flux distribution. If only two outcomes exist, i.e. if the state is only supported by the superposition of an intact and broken configuration, $\Ein=1/2$ implies equal probabilities. The actual state could contain further flux outcomes but as we will see in later sections, the separation determined by $\Ein=1/2$ matches the one from equal weights to numerical precision.

At $m=0$, evaluating Eq.~\eqref{eq:elboson} at the midpoint
gives
\begin{equation}
 \Ein(d,0)=e^{-\mu d/2},\qquad
 \dbreak(0)=\frac{2\ln2}{\mu}.
 \label{eq:massless-field}
\end{equation}
This finite separation describes the overlap of the screening clouds in the exact solution. The massless case is pure screening and does not contain a flux tube, so the resulting separation identifies the half screening point. The spatially resolved profiles show that screening charge accumulates near the external charges, depleting the interior.

\begin{figure}[t]
 \centering
 \includegraphics[width=\columnwidth]{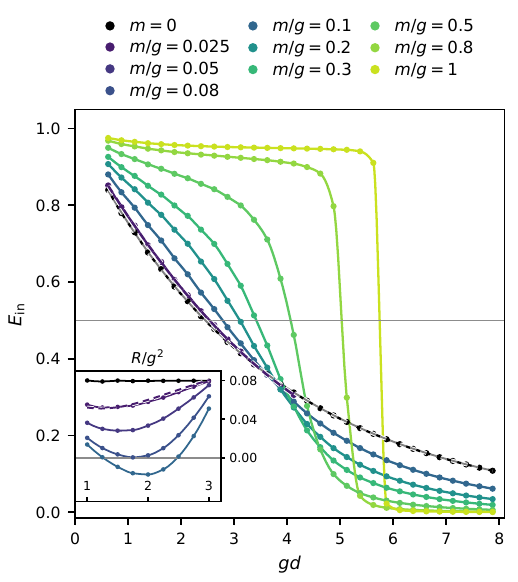}
 \caption{The vacuum subtracted electric field at midpoint between the external charges $\Ein$ for $q=1, \,\theta=0$, on a ring with $N=160$ and $ga=1/8$. The gray dashed curve is the massless free-boson result Eq.~\eqref{eq:massless-field}. The colored dashed curve at $m/g=0.025$ incorporates the first-order
 mass correction. The horizontal line identifies $\Ein=1/2$.
 Increasing $m/g$ yields a convex region in $gd$ that leads to a plateau at $m/g$ followed by a sharp transition.
 The inset shows the curvature $R$ of Eq.~\eqref{eq:curvature} for the light masses. Negative curvature
 indicates the developing concave region and is an imprint of the formation and breaking of the flux tube.}
 \label{fig:center_field}
\end{figure}

The ground-state field changes continuously with separation. At larger
masses, the spectrum resolves an avoided crossing due to mixing between
intact and screened configurations. We characterize
the width $W$ of the crossover by the slope at $\dbreak$
\begin{equation}
 W^{-1}=
 \left|\frac{d\Ein}{d(gd)}\right|_{d=\dbreak}.
 \label{eq:width-definition}
\end{equation}
$W/g$ is a length and $W/(g\dbreak)$ relates it to the length of the string at breaking. The string breaking distance $\dbreak$ is the midpoint where $\Ein=1-Q_{\rm screen}$ half a unit of screening charge has accumulated in the left half of the system. 
With increasing $m/g$ the numerically computed $\Ein$ develops a shoulder followed by a progressively more abrupt fall, as shown in Fig.~\ref{fig:center_field}.

The developing shoulder is visible in the relative curvature,
\begin{equation}
 R(d)=\frac{\Ein''(d)}{\Ein(d)},
 \label{eq:curvature}
\end{equation}
where primes denote derivatives with respect to $d$.

The massless exponential is convex, with $R=\mu^2/4$. Numerically, we find that at any finite mass $R$ is pulled down and $\Ein$ starts developing a concave region above $m/g\sim 0.08$ as shown in the inset of Fig.~\ref{fig:center_field}. The rate change is first caused by the string building up, yielding a slower decay and then by the abrupt flux change at breaking. 

\begin{figure}[t]
 \centering
 \includegraphics[width=\columnwidth]{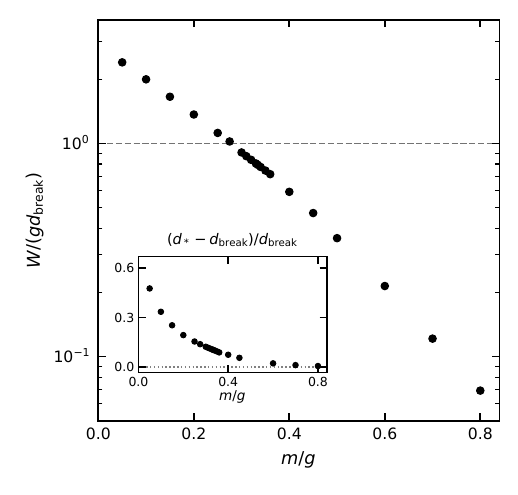}
 \caption{Discharge width and gap minimum as a function of $m/g$.
 Main: width-to-separation ratio $W/(g\dbreak)$.
 Inset: Relative separation of the location of the gap minimum $d_*$ and $\dbreak$.}
 \label{fig:width-gap}
\end{figure}

The width $W$ characterizes how abrupt the transition between intact string and screened configuration happens and is shown in Fig.~\ref{fig:width-gap}. Notably for small masses, the width is larger than the string itself at breaking and the transition is predominantly screening. 
The ratio $W/(g\dbreak)$ decreases smoothly from its massless
value $2/\ln2$ and falls below one near $m/g=0.3$.
The gap minimum also approaches $\dbreak$ as the lowest excitation acquires avoided-crossing character as the gap decreases substantially.

\section{The breaking separation from screened energies}
\label{sec:balance}

In the static case, string breaking can also be understood as a competition between the energy of the intact and broken configuration. The screening clouds contribute to both energies and are dominant when their size is comparable to the length of the string. We construct a model of these energies from three independent inputs: the vacuum response, the meson mass $M_s$ and the static-light pair energy $M_H$. Mass perturbation theory first shows which cloud and bulk contributions this model must distinguish.

\subsection{Mass perturbation and the two meson scales}
\label{sec:mass-corrections}

An applied electric field displaces the mean of the massless boson without changing its fluctuations. We can get the first order correction to the static potential using the Feynman--Hellmann theorem\begin{equation}
 \left.\partial_m V(d,m)\right|_{m=0}
 =\Sigma_0\int dx\,[1-\cos(2\pi S_0(x;d))].
 \label{eq:FH-screening}
\end{equation}

The electric field can be obtained by expanding the screening field $\langle S\rangle=S_0+mS_1+O(m^2)$ and solving Eq.~\eqref{eq:bosonH} to first order (setting $q=1$ throughout)
\begin{gather}
 (-\partial_x^2+\mu^2)S_1(x;d)
 =-2\Sigma_0\sin[2\pi S_0(x;d)],
 \label{eq:firstorder-profile-equation}\\
 S_1(x;d)=-2\Sigma_0\int dy\,G_0(x-y)
                  \sin[2\pi S_0(y;d)].
 \label{eq:firstorder-profile-solution}
\end{gather}
Using the Green function defined in Sec.~\ref{sec:massless-screening} an analytical solution can be obtained. Since it is not insightful for the current discussion, we leave it implicit and note that $S_1(0;d)$ is negative at the massless crossing and hence the correction moves $\dbreak$ to larger separations. 
The first-order field follows the numerical profiles at $m/g=0.025$ and $0.05$ (shown in Fig.~\ref{fig:center_field}). The second derivative which is relevant for $R$ departs sooner and the perturbative curve becomes concave already at $m/g=0.075$, where the all order numerical curve remains convex.

For well-separated external charges, Eq.~\eqref{eq:FH-screening} reduces to
two isolated screening clouds whose mass we determine analytically in the following. The massless profile is $S_0=\tfrac12e^{\mu\xi}$ for $\xi<0$ and $S_0=1-\tfrac12e^{-\mu\xi}$ for $\xi>0$ where $\xi$ denotes the spatial coordinate. The two sides give equal contributions to the condensate integral which is the first order correction to the Hamiltonian. Substituting $u=\pi e^{-\mu|\xi|}$ yields
\begin{equation*}
 I_{\Sigma}=\!\!\int\!\! d\xi\,[1\!-\!\cos(2\pi S_0)]
 \!=\!\frac{2}{\mu}\int_0^\pi\!\!\! du\,\frac{1-\cos u}{u}\equiv\frac{2}{\mu}\operatorname{Cin}(\pi).
 \label{eq:edge-condensate-integral}
\end{equation*}
Each isolated charge contributes $\Sigma_0 I_{\Sigma}$, so the two
screening clouds give $\left.\partial_mM_H\right|_{m=0}=2\Sigma_0 I_{\Sigma}$.
For the vacuum meson, only the quadratic normal-ordered term contributes
to the diagonal one-boson matrix element at this order, giving
$M_s^2=\mu^2+4\pi\Sigma_0m+O(m^2)$~\cite{Dempsey:2022nys}.
The two slopes are therefore
\begin{equation}
 \left.\partial_m M_H\right|_0
 =\frac{2e^\gamma\operatorname{Cin}(\pi)}{\pi},\qquad
 \left.\partial_m M_s\right|_0=e^\gamma.
 \label{eq:MH-Ms-slopes}
\end{equation}
The static-light energy grows slightly faster at small mass because it
samples the full screening cloud, while the meson shift probes a
one-boson excitation of the vacuum.

\begin{figure}[t]
 \centering
 \includegraphics[width=\columnwidth]{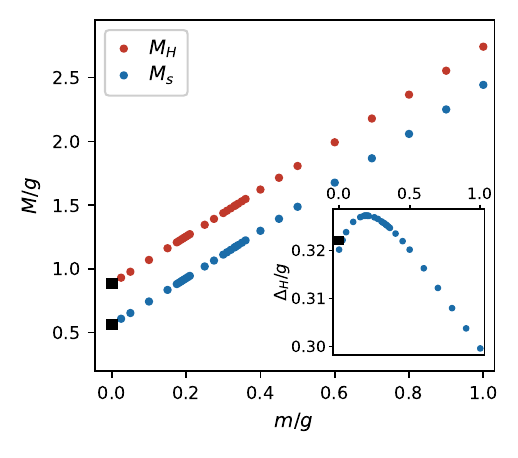}
 \caption{Static-light pair energy $M_H$ and neutral vacuum meson
 mass $M_s$, at $ga=1/8$. The inset shows their difference
 $\Delta_H=M_H-M_s$. Black squares give the exact massless values.
 The initial rise of $\Delta_H$ is fixed by
 Eq.~\eqref{eq:MH-Ms-slopes}, while its shallow maximum near
 $m/g=0.2$ is a numerical result.}
 \label{fig:MH_Ms}
\end{figure}

We denote their difference by $\Delta_H=M_H-M_s$ where the massless difference is $\Delta_H(0)=g(\pi-2)/(2\sqrt\pi)$.
The numerical results in Fig.~\ref{fig:MH_Ms} show a shallow maximum near $m/g=0.2$ followed by a gradual decrease. Their difference therefore retains a mass dependence even where the two energies appear nearly parallel. In particular, $M_s$ interpolates between two slopes since it approaches a slope of 2 in the heavy fermion limit.

\subsection{Vacuum curvature and a finite unit of flux}
\label{sec:vacuum-tension}


The string tension $\sigma$ can be computed from the response of the energy density $\epsilon_{\rm vac}(m,\theta)$ of the  $\theta$-vacuum to a uniform background electric field captured by the topological
susceptibility,
\begin{equation}
 \chi_{\rm top}=\left.\partial_\theta^2
                  \epsilon_{\rm vac}(m,\theta)\right|_{\theta=0}.
 \label{eq:topological-susceptibility}
\end{equation}
At first order in mass, the displaced free vacuum gives $\epsilon_{\rm vac}(m,\theta)-\epsilon_{\rm vac}(m,0)
=m\Sigma_0(1-\cos\theta)+O(m^2)$%
~\cite{Gross:1995bp}.
Since CP acts as $\theta\to-\theta$, the vacuum energy is an even function of $\theta$ and an expansion around $\theta=0$  starts with $\tfrac12\chi_{\rm top}\theta^2$. Writing $F=\theta/(2\pi)$
, the energy density is
\begin{align}
 \epsilon_{\rm vac}(m,2\pi F)-\epsilon_{\rm vac}(m,0)
 &=2\pi^2\chi_{\rm top}F^2+O(F^4),
 \label{eq:vacuum-curvature}\\
 \partial_F\epsilon_{\rm vac}
 &=\frac1L\langle\partial_F H\rangle=g^2\langle E\rangle.
 \label{eq:vacuum-electric-response}
\end{align}
Solving this expansion for $\langle E\rangle$ determines the linear response at weak-field. To relate it to the full $q=1$ external electric field of the string requires an approximation to the functional form of the energy of a chosen branch. The simplest such model is a harmonic approximation of the mass term around one integer $S=n$, together with the exact electric energy:
\begin{equation}
 U_n(S;F)=\frac{g^2}{2}(F-S)^2+\frac K2(S-n)^2.
 \label{eq:harmonic-branch-potential}
\end{equation}
Expanding the cosine in Eq.~\eqref{eq:bosonH} gives
$K=4\pi^2m\Sigma_0$. We allow an effective $K$ determined by the
interacting vacuum response. The electric term favors $S=F$,
while the restoring term favors $S=n$. Minimizing
$U_n$ with respect to $S$ determines its mean $\overline S_n$ and the branch energy
density $\bar\epsilon_n^{\rm h}$:
\begin{align}
 \overline S_n-n&=\frac{g^2}{g^2+K}(F-n),
 \label{eq:harmonic-polarization}\\
 \bar\epsilon_n^{\rm h}(F)-\epsilon_{\rm ref}^{\rm h}
 &=\sigma_{\rm h}(F-n)^2,\qquad
 \sigma_{\rm h}=\frac{g^2K}{2(g^2+K)}.
 \label{eq:polarized-branch-energy}
\end{align}
The reference density is independent of $F$ and $n$. Displacement of a
quadratic quantum field also leaves its fluctuation frequencies
unchanged, so these expressions give its source-dependent energy as
well as the minimum of the classical functional.

Within this model, the coefficient of the weak-field response also
determines the difference between neighboring branches at $F=1$:
\begin{equation}
 \bar\epsilon_0^{\rm h}(1)-\bar\epsilon_1^{\rm h}(1)
 =\sigma_{\rm h}=2\pi^2\chi_{\rm top}^{\rm h}.
 \label{eq:harmonic-tension}
\end{equation}
Matching the vacuum curvature fixes the bulk coefficient of the model,
\begin{equation}
 \sigma=2\pi^2\chi_{\rm top}.
 \label{eq:sigma-chitop}
\end{equation}
At $F=1$, the $n=0$ continuation leaves an electric field $2\sigma/g^2$
and with energy cost $\sigma$ per unit length. The neighboring $n=1$ continuation
has $\overline S_1=1$, which cancels the imposed field and removes
this bulk energy. For a finite string, the screened interior must
still connect to the exterior vacuum. The screening clouds at these
two boundaries have the finite total energy $M_H$ at large separation.
Thus $\sigma$ is the modeled energy per unit length of the intact
interior relative to the fully screened interior.

This continuation is justified as a local expansion when $|\overline S_n-n|$ is small. Screening is suppressed at large $m/g$, so this can hold even at the full external field of the string. Light fermions can screen much of that field within the same parabola, taking the displacement beyond the range of the local expansion. The omitted cosine terms then contribute already at first order in $m$. The curvature alone consequently does not determine the finite-field energy. Below, we test the continuation through its predicted string crossing. After allowing complete relaxation between branches, the asymptotic tension instead follows the periodic ground-state energy~\cite{Gross:1995bp,Adam:1996rd}:
\begin{equation}
 \sigma_q^\infty=
 \epsilon_{\rm vac}(m,\theta+2\pi q)
 -\epsilon_{\rm vac}(m,\theta),
 \label{eq:sigmaq}
\end{equation} which vanishes for integer $q$ by the periodicity of the vacuum energy.

The meson mass provides a second way to fix the quadratic model. Keeping the kinetic term $\Pi_S^2/(2\pi)$ and gradient term $\pi(\partial_xS)^2/2$ of Eq.~\eqref{eq:bosonH}, a potential of curvature $g^2+K$ gives $M_s^2=(g^2+K)/\pi$. Eliminating $K$ then yields
\begin{equation}
 \sigma_M=\frac{g^2}{2}\left(1-\frac{g^2}{\pi M_s^2}\right).
 \label{eq:mass-curvature-check}
\end{equation} 
This relation follows when the response is that of a single free massive field. In the interacting vacuum, the response also depends on the matrix elements coupling the vacuum to its excitations which generally does not have to be determined by the lowest excitation. However, we find excellent numerical agreement of $\sigma_M$ with $\sigma$ in the $\theta$-vacuum as shown in Fig.~\ref{fig:energy_balance}.

\subsection{Screening clouds and configuration energies}
\label{sec:screened-energies}

To obtain the energy of a finite string, replace the uniform electric field $F$ in $U_0$ by the source profile $f_d(x)$ and restore the gradient energy of Eq.~\eqref{eq:bosonH}. The gradient term accounts for the screening clouds where the interior joins the exterior vacuum. We fix the uniform response by $\sigma$ and the cloud decay length by $M_s^{-1}$.  The minimizing profile satisfies
\begin{equation}
 (-\partial_x^2+M_s^2)S_I=
 \left(1-\frac{2\sigma}{g^2}\right)M_s^2 f_d
 \label{eq:intact-profile-equation}
\end{equation} which has the same spatial form as Eq.~\eqref{eq:massless-profile}, with $\mu$ replaced by $M_s$ and amplitude $1-2\sigma/g^2$. In particular,
\begin{equation}
 E_{\rm in}^{I}(d)=\frac{2\sigma}{g^2}
 +\left(1-\frac{2\sigma}{g^2}\right)e^{-M_sd/2}.
 \label{eq:intact-configuration-field}
\end{equation} 
Screening lowers the interior field of this intact configuration, but leaves the bulk value $2\sigma/g^2$ at large separation. Both this field and the associated energy follow from the same response functional.

We denote the energy of this intact profile by $\energy_I(d)$.
At the minimum, the profile equation relates the gradient and
restoring energies to the source term. Their sum reduces to
\begin{align}
 \energy_I(d)&=\frac{g^2}{2}\left[d-\int_{-d/2}^{d/2}S_I(x;d)\,dx\right]
 \nonumber\\
 &=\sigma d+c_I(1-e^{-M_sd}),
 \label{eq:EIform}\\
 c_I&=\frac{g^2/2-\sigma}{M_s}.
 \label{eq:cIderived}
\end{align}
The extensive term is the energy per unit length required for the string.
The remaining energy comes from the two boundaries and includes the
overlap of their screening clouds. The coefficient is determined by
$\partial_d\energy_I(0^+)=g^2/2$, while at large separation the slope
approaches $\sigma$. Varying $d$ moves the external charges, so
$\partial_d\energy_I$ is the force required to separate them. The midpoint electric field in Eq.~\eqref{eq:intact-configuration-field}
is a different evaluation of the same spatial solution.

In the screened configuration the external charges bind two static-light
mesons of total energy $M_H$. Their residual interaction decays over
the meson length $M_s^{-1}$ if the lightest neutral mode controls its
tail. We approximate this attraction by a single exponential and extend it through the region of overlapping clouds. Imposing zero energy for coincident sources fixes its amplitude, giving the modeled
screened energy
\begin{equation}
 \energy_B(d)=M_H(1-e^{-M_sd}).
 \label{eq:EBform}
\end{equation}
At $m=0$, Eq.~\eqref{eq:EBform}
is the exact potential. At finite mass, it is an interpolation which is only valid near the crossing with $\energy_I$.

\subsection{The energy crossing and string breaking}
\label{sec:energy-balance}

The intact and screened energies have a nonzero crossing $d_x$ when
\begin{equation}
 \sigma d_x=(M_H-c_I)(1-e^{-M_sd_x}).
 \label{eq:energybalance}
\end{equation}
For $\sigma>0$ and $M_sM_H>g^2/2$, the screened energy initially rises
faster, but the intact energy eventually exceeds its plateau.
Defining $\beta=M_s(M_H-c_I)/\sigma>1$, the unique positive root is
\begin{equation}
 d_x=\frac{\beta+\mathrm W_0(-\beta e^{-\beta})}{M_s},
 \label{eq:dbreaksolved}
\end{equation}
where $\mathrm W_0$ is the principal Lambert function, defined by
$\mathrm W_0(z)e^{\mathrm W_0(z)}=z$. The other real branch gives the solution $d=0$.

\begin{figure}[t]
 \centering
 \includegraphics[width=\columnwidth]{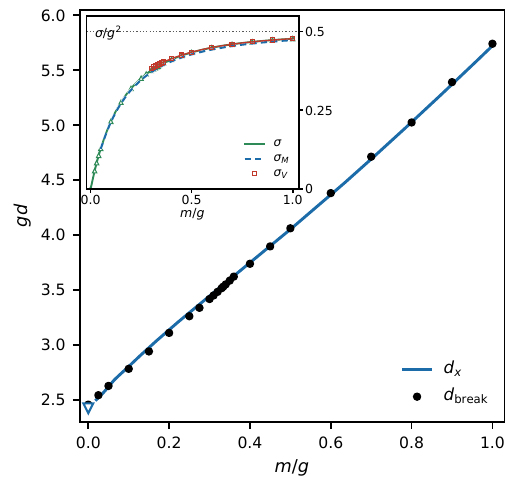}
 \caption{Comparison of $\dbreak$ (points) and the screened
 energy crossing $d_x$ (curve), Eq.~\eqref{eq:dbreaksolved}, at
 $ga=1/8$. The vacuum susceptibility, meson mass and static-light plateau fix the curve independently. The open triangle is the analytic $m\to0^+$ limit of $d_x$, distinct from $\dbreak(0)$.
 The inset compares $\sigma=2\pi^2\chi_{\rm top}$ with the mass-based prediction $\sigma_M$, Eq.~\eqref{eq:mass-curvature-check} and the
 slope $\sigma_V$ of the approximately linear part of the static
 potential. The latter is shown only for $m/g\geq0.3$, where this
 linear part can be resolved.}
 \label{fig:energy_balance}
\end{figure}

Figure~\ref{fig:energy_balance} compares this energy crossing with
$\dbreak$ obtained from the midpoint electric field. The agreement is excellent for all masses considered. Note that the inputs are determined independently of that separation: $M_s$ and $\sigma$ come from vacuum calculations and $M_H$ from the static potential at large separation. The agreement with $\dbreak$ connects the local loss of flux to the energetic competition between an intact string with the screened configuration.

Where the potential contains a resolved linear interval, its slope provides a separate test of the bulk coefficient, shown in
the inset of Fig.~\ref{fig:energy_balance}. At small mass screening obscures the linear regime,
while the string tension from the vacuum topological susceptibility remains well defined. The mass relation
in Eq.~\eqref{eq:mass-curvature-check} gives an approximate analytical expression for $\sigma$, also shown in the inset. 

The limits show what the screening terms contribute. When
$M_sd_x\gg1$, the crossing approaches $(M_H-c_I)/\sigma$.
For weak polarization, $\sigma\to g^2/2$ and $c_I/M_H\to0$, recovering
$M_H/\sigma$. In the massless limit the opposite occurs:
$c_I=M_H(0)$ and both energy functions reduce to the exact potential
in Eq.~\eqref{eq:massless-potential} at every separation. Their equality
then selects no crossing, although $\dbreak$ remains
finite.

Approaching from positive mass still gives a definite limit. Using
Eq.~\eqref{eq:MH-Ms-slopes} and
$\left.\partial_m\sigma\right|_0=2\pi^2\Sigma_0$,
\begin{equation}
 \beta_0=\lim_{m\to0^+}\beta
 =\frac32+\frac{2\operatorname{Cin}(\pi)}{\pi^2}.
 \label{eq:beta0}
\end{equation}
The corresponding limit of Eq.~\eqref{eq:dbreaksolved} is
$gd_x\simeq2.42$, slightly below the exact half-field value
$g\dbreak(0)=2\sqrt\pi\ln2\simeq2.46$. The screened energy balance is therefore
an approximation to the discharge location, while reproducing the
energy exactly at zero mass.


\begin{figure*}[t]
 \centering
 \includegraphics[width=.48\textwidth]{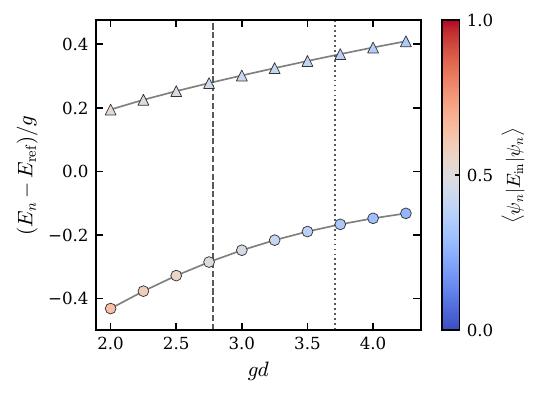}\hfill
 \includegraphics[width=.48\textwidth]{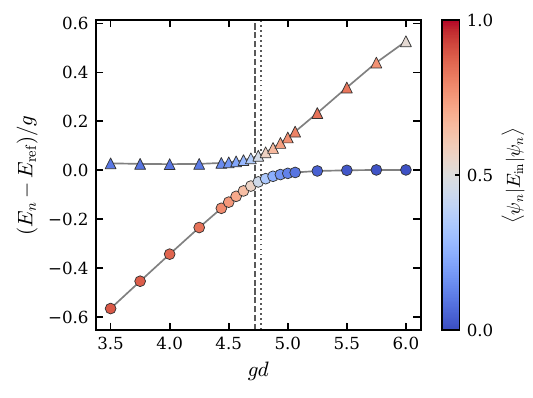}
 \par\makebox[.48\textwidth]{(a)}\hfill\makebox[.48\textwidth]{(b)}
 \caption{Ground (circles) and first-excited (triangles) energy
 levels, colored by their vacuum-subtracted midpoint fields, for
 unit sources at $\theta=0$.
 (a) $m/g=0.1$, $N=160$, $ga=1/8$.
 (b) $m/g=0.7$, $N=400$, $ga=1/32$.
 Both levels at each mass use the same constant energy reference,
 $\energy_{\rm ref}=(\overline{\energy}_0+
 \overline{\energy}_1)/2$ evaluated at $\dbreak$.
 Overbars denote averages over adjacent opposite source placements.
 The plotted energy symbols $E_n$ and $E_{\rm ref}$ denote
 $\energy_n$ and $\energy_{\rm ref}$ in the text.
 The dashed vertical line locates $\dbreak$ and the dotted line
 $d_*$. At light mass both states lose flux. At larger mass their
 opposite field changes identify an avoided crossing between string
 and screened-pair configurations.}
 \label{fig:string-levels}
\end{figure*}

Fig.~\ref{fig:string-levels} shows the interior electric field of the ground and first excited state. At small mass (a), the electric field has the same orientation as the ground state and is a bound meson while for the larger mass it is oppositely oriented and corresponds to the crossing partner (b).

\subsection{Anti-aligned background and the vacuum limit}
\label{sec:background-field-dial}

An opposing background electric field reduces the energy gained by discharging
the string. This can be seen from the quadratic branch description of
Sec.~\ref{sec:vacuum-tension} for $F=\theta/(2\pi)$.
The intact interior has displacement $F+1$ from its branch
minimum. A unit of screening changes that displacement to $F$.
Subtracting the two bulk energy densities gives
\begin{equation}
 \sigma(\theta)=\sigma\bigl[(F+1)^2-F^2\bigr]
 =\sigma\left(1+\frac{\theta}{\pi}\right).
 \label{eq:sigmatheta}
\end{equation}
As $\theta$ approaches $-\pi$, a longer interval is needed
for this bulk energy gain to compensate the finite cost of
the screening charges. The configuration crossing
therefore moves to larger separations. The linear dependence on $\theta$ in
Eq.~\eqref{eq:sigmatheta} uses the harmonic continuation.
For $m>m_c$, the ground state at $-\pi$ is degenerate~\cite{Coleman:1976uz,Byrnes:2002gj}.
At this angle, the interval with background field $\theta=+\pi$ can support
the polarization opposite to the exterior at $\theta=-\pi$. Increasing $d$ then enlarges a domain of the competing vacuum with the sources at its interfaces. The string construction then
approaches vacuum coexistence when both $d$ and
the exterior length $L-d$ grow.

Numerically we find that the $m/g=0.5$ spectrum follows this trend: the minimum gap
decreases and its location $d_*$ moves to larger separation as the background approaches $-\pi$.
At the largest separations, the exterior is comparable in length to the string and the mass gap tends to zero as $\theta\to-\pi.$ In contrast, the gap remains finite below the critical point for $m/g=0.25$ even at $\theta=\pi$.

\section{Flux-resolved entanglement and pair production}
\label{sec:string-branch-qis}
\label{sec:qis}

The electric field describes how much flux is discharged. The
reduced state of a spatial interval reveals which correlations
accompany that change. Resolving its boundary electric fluxes
separates uncertainty between screening alternatives from
entanglement within each alternative.
As we will show in this section, near breaking, the coherently selected intact and screened sectors
reveal a nearly maximally entangled bit shared by the interval and
its complement. Its approximate Bell factorization separates the
screening information from residual correlations after local encoding.
The associated binary uncertainty supplies most of the
peak of the entanglement entropy at breaking, while entanglement within each branch also increases.
We quantify this factor and its vacuum counterpart, then compare
string and vacuum intervals through their capacity of entanglement.

\subsection{Flux sectors of a spatial interval}
We retain the electric fields on both boundaries of an interval
$A$ as part of its gauge-invariant algebra, the electric-center
choice. Operators confined
to this algebra preserve the bounding fluxes, which therefore
specify separate sectors~\cite{Casini:2013rba,Feldman:2024qif}.
For a pure state of the full system, the reduced density
matrix of $A$ has the direct-sum form
\begin{equation}
 \rho_A=\bigoplus_k\rho_A^{(k)}
       =\bigoplus_k w_k\sigma_k,\qquad
 w_k=\operatorname{tr}\rho_A^{(k)},\quad
 \operatorname{tr}\sigma_k=1 .
 \label{eq:qis-blocks}
\end{equation}
Here $k$ specifies the total bounding electric fields
$E_L,E_R$, including the imposed background and source fields
in Eq.~\eqref{eq:fielddef}. For an interval containing no static
source, Gauss's law fixes its charge to $Q_A=E_R-E_L$. Fixing
the charge alone leaves the
common boundary flux undetermined, so the pair of fluxes
contains information beyond the enclosed charge.

We denote the eigenvalues of $\sigma_k$ by $\lambda_{ki}$.
The eigenvalues of $\rho_A$ are $w_k\lambda_{ki}$.
Separating their logarithms in the von Neumann entropy
$S_A=-\operatorname{tr}\rho_A\ln\rho_A$ gives
\begin{align}
 S_A&=\Sconf+\Ssec,\qquad S_k=-\operatorname{tr}\sigma_k\ln\sigma_k,
 \label{eq:entropysplit}\\
 \Sconf&=-\sum_k w_k\ln w_k,\qquad
 \Ssec=\sum_k w_kS_k .
 \nonumber
\end{align}
We use natural logarithms for these entropies. $\Sconf$ is the
uncertainty in the bounding fluxes, whereas $\Ssec$ is the
average entanglement remaining after those fluxes are
specified.\footnote{We use configurational entropy for
the Shannon entropy of the flux probabilities. In the
symmetry-resolved entanglement literature that name also
denotes the complementary, within-sector term.}
Gauge-invariant operations restricted to the two regions
can distill the latter contribution~\cite{VanAcoleyen:2015ccp}.
This restriction concerns local operations across a fixed
partition. The Hamiltonian can still create pairs and
transfer charge through its boundaries.

For the string, we take $A$ to contain four staggered sites
at the midpoint, on a ring with $N=160$ and $ga=1/8$.
At finite spacing, the interval boundaries can lie on either parity
of the staggered lattice. An overbar denotes the average of scalar
quantities evaluated separately for these two boundary alignments.
All flux sectors contribute to the total quantities.

\subsection{Entanglement within the string branches}
\label{sec:branch-projected-qis}

Inside a $q=1$ string at $\theta=0$, the intact branch $I$ has
total flux one on both boundaries of $A$ and the screened branch
$B$ has flux zero on both boundaries of the same interval.
Their probabilities are $w_I$ and $w_B$. Conditioning on either
flux defines a conditional, flux-resolved state, $\sigma_I$ or $\sigma_B$, containing
all compatible matter configurations. These projections distinguish
intact and screened flux in the string interior without specifying
complete intact- or broken-string wavefunctions.

In the heavy-mass limit, away from the crossing, one branch
dominates and the uncertainty between intact and screened flux
vanishes. This agrees with the results of ~\cite{Amorosso:2024glf}  who showed that a definite Abelian flux tube has no intrinsic color entanglement in pure
$(1+1)$-dimensional gauge theory. This is the
pure-gauge limit of the heavy string away from breaking.
At finite mass, dynamical fluctuations populate additional
flux sectors and branch competition produces an entropy
peak near the crossing.

\begin{figure*}[t]
 \centering
 \includegraphics[width=\textwidth]{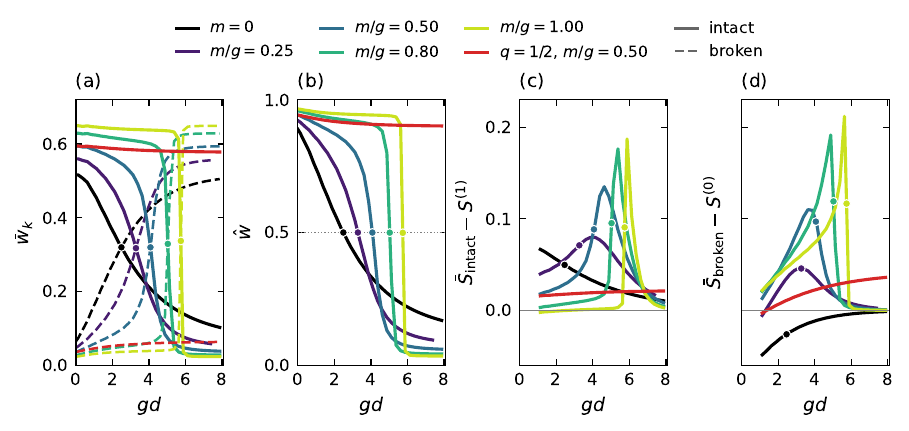}
 \caption{String branch probabilities and flux-resolved entropies
 for a four-site central interval, $N=160$, $ga=1/8$ and
 $\theta=0$. Colors specify the masses for $q=1$ and the
 $q=1/2$ comparison at $m/g=0.5$.
 (a) Absolute probabilities of intact flux one (solid) and
 screened flux zero (dashed) for the unit strings.
 (b) Their conditional ratio, Eq.~\eqref{eq:qis-conditional-weight}.
 Dots in (a,b) locate equal probabilities.
 (c,d) Entropies of the normalized intact and broken blocks,
 relative to the $\theta=0$ vacuum blocks of total flux
 one and zero, denoted $S^{(1)}$ and $S^{(0)}$.
 Dots in (c,d) locate $\dbreak$ from the field.
 The half-charge curves use the same reference blocks,
 although their sourced bounding fields are $1/2$ and $-1/2$.
 Entropy curves begin where the interval lies entirely
 between the sources. Entropies are evaluated before
 averaging over the two placements.}
 \label{fig:branch-qis-allmass}
\end{figure*}

Figure~\ref{fig:branch-qis-allmass}(a) shows the exchange
of these two probabilities as the source separation grows.
Notably, the weights of the unbreakable $q=1/2$ string do not exchange. To display the
exchange within the selected pair, panel (b) uses
\begin{equation}
 \hat w=\frac{\bar w_I}{\bar w_I+\bar w_B} .
 \label{eq:qis-conditional-weight}
\end{equation}
At their crossing, $\bar w_I=\bar w_B$ and $\hat w=1/2$.
The equal-weight crossing matches the independently determined
$\dbreak$ within our resolution, linking the
$\Ein$ criterion to the exchange of flux probabilities.
The rise in $\Sconf$ supplies
most of the entropy peak and the smaller rise in $\Ssec$
resolves entanglement within the sectors.

The other sectors also reorganize at the crossing. For the crossings through $m/g=0.5$, the
distribution is approximately invariant under
$(E_L,E_R)\mapsto(1-E_R,1-E_L)$, including the less probable
sectors. This reflection exchanges intact and screened flux
and sends the mean bounding field $E$ to $1-E$.
The reflection symmetry at breaking is an observed property of the string distribution and even holds when a background electric field is switched on with $\theta$ where CP is generally broken.

The flux-resolved entropies characterize the flux entanglement on the flux tube and between the produced mesons and is shown in panel (c) and (d) of  Fig.~\ref{fig:branch-qis-allmass}. The appropriate vacuum reference are the vacuum sectors carrying the same total bounding flux which is
flux one for the intact sector and flux zero for the
broken sector. At finite mass, both flux-resolved entropies
develop a maximum near breaking.
The flux-resolved peaks therefore establish that screening changes
the internal states, beyond redistributing their probabilities. The broken-sector entropy reaches
its maximum by the crossing, while the intact-sector
entropy continues to rise beyond it. The narrowing at
large mass limits the determination of the peak positions.
Neither the massless string nor the half-charge string
has the corresponding pair of flux-resolved entropy peaks.

\subsection{Maximal entanglement and the screening bit}
\label{sec:qis-thermality}

Screening lowers the interior flux by one unit. A central interval
and its complement therefore carry shared records of intact or
screened flux. In the electric-center description these records are
orthogonal on both sides of the partition. The operators $P_{I,A}$ and
$P_{B,A}$ project onto total bounding flux one and zero, respectively.
Every observable $O_A$ in the electric-center algebra of $A$
preserves these fluxes, so $P_{I,A}O_AP_{B,A}=0$.
Measurements confined to $A$ distinguish the branches but cannot recover their relative phase. The complement retains the flux record that makes their interference inaccessible in $A$.

For this interval in the ground state $|\Psi\rangle$, coherently retaining the two sectors gives
\begin{equation}
 \begin{aligned}
 |\Psi_{IB}\rangle&=
 \frac{(P_{I,A}+P_{B,A})|\Psi\rangle}{\sqrt{w_{IB}}},\\
 w_{IB}&=w_I+w_B,\qquad p=\frac{w_I}{w_{IB}}.
 \end{aligned}
 \label{eq:qis-selected-pure-state}
\end{equation}
The projection preserves the relative phase and retains all matter
configurations compatible with the selected fluxes. The total
probability of this selection is $w_{IB}$ and $p$ is the intact
probability within it. Orthogonal records on both sides give the entropy of
a biorthogonal superposition%
~\cite{Linden:2005xuc},
\begin{equation}
 S_{A\mid I,B}=h_2(p)+pS_I+(1-p)S_B,
 \label{eq:qis-branch-maximal-entropy}
\end{equation}
where $h_2(p)=-p\ln p-(1-p)\ln(1-p)$. At equal probabilities,
$h_2(1/2)=\ln2$ exactly, regardless of the internal spectra.
This gives a binary realization of Kharzeev's MEL
mechanism~\cite{Kharzeev:2026jkq}: the full selected
state is coherent, while either side has maximal uncertainty about
the shared flux. Flux projection makes the string--meson
interpretation of Ref.~\cite{Grieninger:2025rdi} explicit
without specifying complete wavefunctions for those configurations.

The probability balance also has an interference interpretation. On the
selected subspace, the relative phase flip
$Z_{IB}=P_{I,A}-P_{B,A}$ gives
\begin{equation}
 \langle\Psi_{IB}|Z_{IB}|\Psi_{IB}\rangle=2p-1.
 \label{eq:qis-phase-flip}
\end{equation}
At equal probabilities this operation makes the selected global state
orthogonal to its original state, while leaving the flux-diagonal
reduced state of $A$ unchanged. This characterizes coherent branch
balance, including the smooth massless screening crossover.

A Bell factor requires the binary degree of freedom to separate
from those remaining correlations. Encode intact and screened flux
as $0$ and $1$ on each side and align the flux-resolved Schmidt bases
by local isometries within their supports. We compare the encoded
state with the reference
\begin{equation}
 \begin{aligned}
 |\Psi_{\rm ref}\rangle&=|\Phi_\phi\rangle\otimes|\chi\rangle,\\
 |\Phi_\phi\rangle&=
 \frac{|0\rangle_A|0\rangle_{\bar A}
       +e^{i\phi}|1\rangle_A|1\rangle_{\bar A}}{\sqrt2}.
 \end{aligned}
 \label{eq:qis-bell-reference}
\end{equation}
Here $|\Phi_\phi\rangle$ is an equal-amplitude Bell pair with relative
phase $\phi$. The normalized pure state $|\chi\rangle$ describes the
remaining degrees of freedom on both sides and may itself be
entangled across the partition. The tensor product requires the same
$|\chi\rangle$ for either flux, making the screening bit independent
of those remaining correlations. Optimizing over $|\chi\rangle$
and the branch-dependent local isometries gives the squared fidelity
\begin{equation}
 \begin{aligned}
 F_{\rm Bell}&=\frac12+\sqrt{p(1-p)}\,\beta,\\
 \beta&=\sum_i\sqrt{\lambda_{Ii}\lambda_{Bi}}.
 \end{aligned}
 \label{eq:qis-bell-fidelity}
\end{equation}
Here the normalized flux-resolved eigenvalues are sorted in decreasing
order. Appendix~\ref{app:qis-bell} derives
this result and bounds $\beta$ from the leading eigenvalues. Equal probabilities maximize the uncertainty of the shared flux record. Identical flux-resolved spectra additionally permit the same
residual state for both alternatives after local encoding. An exact Bell factor therefore requires $p=1/2$ and $\beta=1$; the entropy contribution $\ln2$ alone does not establish this independence.

Near breaking at $m/g=0.5$, the flux-resolved spectra establish an approximate Bell factor, with fidelity bounds given in
Appendix~\ref{app:qis-bell}. Figure~\ref{fig:qis-bell-fidelity} shows the separation dependence
through bounds on the fidelity. The peak persists as the interval is enlarged to twelve sites.
Both flux-resolved spectra are dominated by a single eigenvalue and remain similar away from breaking. The approach to maximal entanglement therefore comes from balancing the probabilities of intact and screened flux. This also holds at lighter masses where the flux-resolved spectra also permit a nearly independent bit. 

\begin{figure}[t]
 \centering
 \includegraphics[width=\columnwidth]{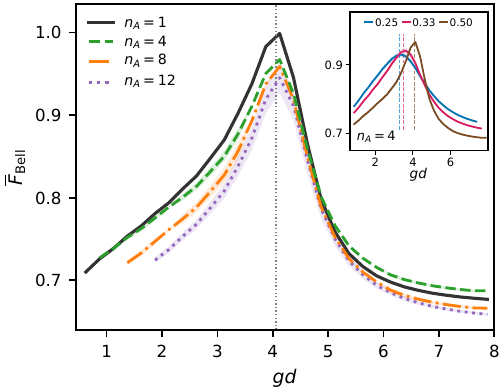}
 \caption{Bounds on the flux-resolved squared Bell fidelity
 for intact and screened flux in a central interval.
 Main: $m/g=0.5$ with interval size $n_A$ in staggered sites.
 The one-site curve is exact. For larger intervals, lines follow
 the midpoints of the bounds and shaded bands span their full range,
 using flux-resolved
 entropies and, where available, leading eigenvalues
 (Appendix~\ref{app:qis-bell}).
 For $n_A>1$, the bar denotes averaging the bounds after evaluating
 them at each boundary alignment. The dotted vertical line marks $\dbreak$.
 Inset: four-site intervals at the masses $m/g$ given in the legend.
 The dashed vertical lines mark the corresponding breaking distances.
 Parameters are $N=160$, $ga=1/8$, $q=1$ and $\theta=0$.}
 \label{fig:qis-bell-fidelity}
\end{figure}

The resulting bit records whether dynamical charges have screened
the interior flux and nearly separates from the remaining correlations.
Its two qubits belong to the string interior and its complement.
Both external sources lie in the complement, so this partition
does not assign one qubit to each screening charge. The local
isometries characterize the factorization in the electric-center
embedding. Gauge-invariant operations confined to either side preserve the flux
records, so they cannot distill the binary contribution%
~\cite{VanAcoleyen:2015ccp}.
In the full spatial entropy, the selected binary term is $w_{IB} h_2(p)$.

\subsection{An emergent temperature for configuration probabilities}
\label{sec:qis-probability-response}

Quantum mixing sets an energy scale for the configuration probabilities
when the two competing configurations form an isolated low-energy
doublet. For string--meson mixing~\cite{Bali:2005fu}, take a
locally fixed orthonormal configuration basis with diagonal bias
$\varepsilon=\energy_B-\energy_I$ and effective off-diagonal
coupling $-\Gamma$, with $\Gamma>0$ locally constant. In this basis,
$H_2=\overline{\energy}\,\mathbf1-\varepsilon\tau_z/2-\Gamma\tau_x$,
where $\overline{\energy}=(\energy_I+\energy_B)/2$, $\tau_x,\tau_z$
are Pauli matrices and the intact configuration has $\tau_z=+1$.
The basis states describe the dressed configurations, including
polarization. Below, $p_I,p_B$ are their probabilities in the
effective ground state, distinct from probabilities conditioned on
two boundary-flux sectors. Eliminating intermediate configurations
contributes to $\Gamma$~\cite{RevModPhys.36.1076}, allowing
effective tunnelling even when direct hopping between the selected
flux blocks vanishes. The ground-state configuration probabilities obey
\begin{align}
 p_I&=\frac12\left(1+\frac{\varepsilon}
                   {\sqrt{\varepsilon^2+4\Gamma^2}}\right),
 \qquad p_B=1-p_I,
 \nonumber\\
 \ln\frac{p_I}{p_B}
 &=2\operatorname{arsinh}\frac{\varepsilon}{2\Gamma}
   =\frac{\varepsilon}{T_{\rm prob}} .
 \label{eq:qis-probability-temperature}
\end{align}
Matching the odds to Boltzmann weights for the diagonal energies
defines $T_{\rm prob}=\varepsilon/
[2\operatorname{arsinh}(\varepsilon/2\Gamma)]$.
At equal diagonal energies the weights are equal for every positive Boltzmann
temperature; their response to a bias fixes
$T_{\rm prob}\to\Gamma=\Delta(0)/2$, where
$\Delta(\varepsilon)=\sqrt{\varepsilon^2+4\Gamma^2}$ is the
partner gap. This emergent scale gives the Boltzmann form of the
configuration odds in a zero-temperature ground state and varies
with bias. The temperature
inferred from thermal-state comparisons in
Ref.~\cite{Grieninger:2025rdi} instead concerns the full
reduced state. Equation~\eqref{eq:qis-probability-temperature}
determines only the configuration odds.

For $m>m_c$, the vacuum at
$\theta=\pi$  has equal configuration probabilities between configurations with opposite electric field and
$T_{\rm prob}\to\Gamma_L$, half the doublet splitting.

\subsection{Vacuum branches and nonstabilizerness}
\label{sec:vacuum-branch-qis}

The vacuum provides a second test of changes inside the
flux branches. We now take $A$ to be half space i.e. half of the ring.
The two selected neutral sectors carry total bounding flux
$\theta/(2\pi)$ and $\theta/(2\pi)-1$, becoming $+1/2$ and
$-1/2$ at $\theta=\pi$. Specifying these boundary fields leaves
the interior fluctuations unresolved. The orthogonal flux records
give the same phase restriction and binary entropy bound as in
the string (Sec.~\ref{sec:qis-thermality}).

The same Bell-fidelity peak at the transition as observed for the string in Sec.~\ref{sec:qis-thermality} is found for all three masses considered.
At $m/g=0.5$ and $\theta=\pi$, the flux-resolved spectra establish
an approximately independent, nearly maximally entangled flux bit shared by the two halves. At the
lighter masses studied, the behavior is qualitatively similar but the Bell fidelity is reduced, similar to what we observed for the string at lighter mass. 

The flux-resolved spectra also quantify nonstabilizerness,
the quantum resource beyond stabilizer states and
Clifford operations. A pure stabilizer state's reduced
spectrum is flat with rank a power of two. Departures from
this form imply nonlocal magic that cannot be removed
by basis changes confined to either side of the
partition~\cite{Tirrito:2023fnw,Ebner:2026kwq}.
For each normalized flux block we form the canonical
Schmidt state with amplitudes $\sqrt{\lambda_{ki}}$.
The corresponding order-two stabilizer R\'enyi entropy is denoted
$\mathcal M_{2,k}$~\cite{Leone:2021rzd}.
This canonical value is an upper bound on the minimum
under local basis changes. 

\begin{figure*}[t]
 \centering
 \includegraphics[width=\textwidth]{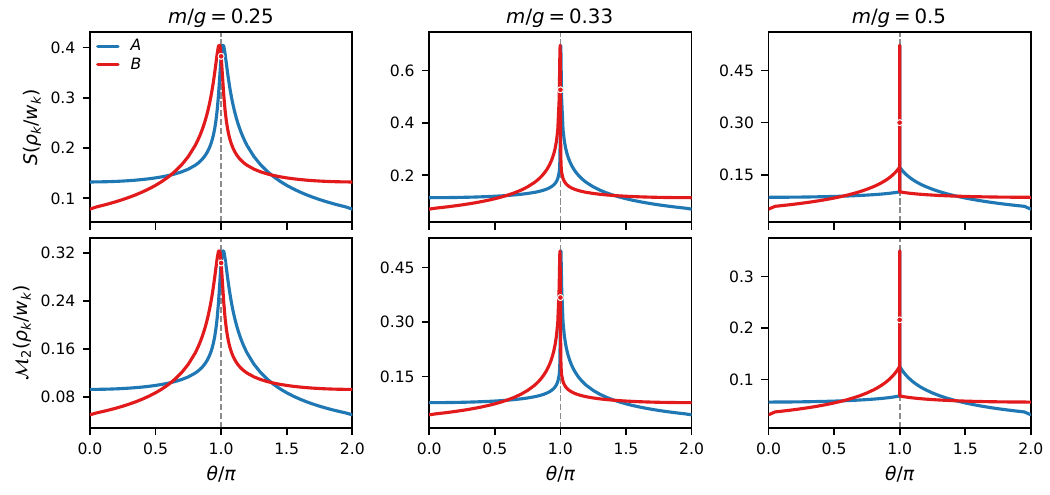}
 \caption{Vacuum information within the two neutral flux
 sectors, for a half-space interval with $N=320$, $ga=1/8$ and $L_{\max}=5$.
 The branches $A$ and $B$ carry total bounding flux $\theta/(2\pi)$ and $\theta/(2\pi)-1$, respectively.
 Upper row: entropy of each normalized block.
 Lower row: canonical stabilizer R\'enyi entropy
 $\mathcal M_{2,k}$.
 The plotted $\rho_k/w_k$ is $\sigma_k$ in
 Eq.~\eqref{eq:qis-blocks}.
 CP averaging exchanges the branch indices as it reflects
 $\theta$ about $\pi$. Dots give the values at $\pi$.
 The corresponding absolute weights are shown in
 Fig.~\ref{fig:qis-vacuum-weights}.
 The fine scan at $m/g=0.5$ resolves changes much narrower
 than the displayed angular scale. Lines interpolate
 the available angles.}
 \label{fig:vac-branch-qis}
\end{figure*}

Figure~\ref{fig:vac-branch-qis} shows that the normalized
branch entropies and their $\mathcal M_{2,k}$ both increase
near the vacuum exchange. CP averaging makes the two curves
equal at $\theta=\pi$.
The individual curves reach their largest values
on opposite sides of the transition, similarly as in case of the string at breaking. 
These increase of the quantities toward the transition identifies changes within the vacuum
branches, paralleling the flux-resolved entropy peaks
of the string.

\begin{figure*}[t]
 \centering
 \includegraphics[width=\textwidth]{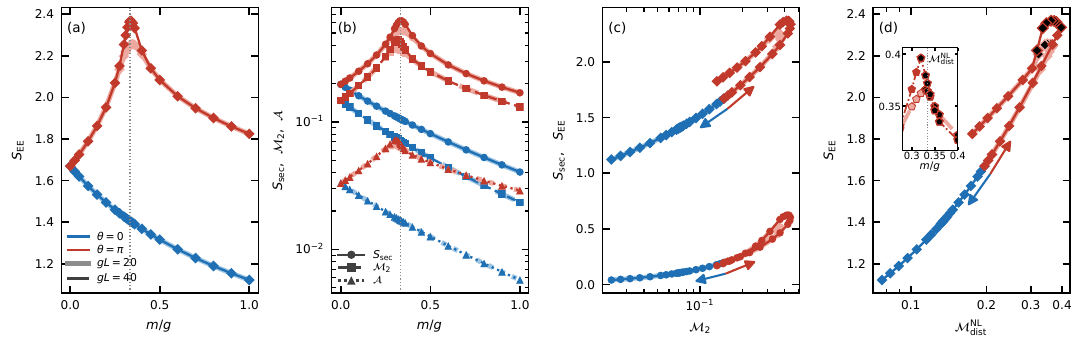}
 \caption{Half-ring vacuum entanglement and flux-resolved
 nonstabilizerness at $\theta=0$ (blue) and $\pi$ (red),
 with $ga=1/8$. Thick pale curves have $gL=20$ and thin
 curves have $gL=40$.
 (a) Total entropy, denoted $S_{\rm EE}$.
 (b) Within-sector entropy $\Ssec$, canonical stabilizer
 R\'enyi entropy $\mathcal M_2$ and antiflatness $\mathcal A$.
 (c) Total and within-sector entropies against $\mathcal M_2$.
 (d) Total entropy against $\mathcal M_{\rm dist}^{\rm NL}$,
 with the $\theta=\pi$ mass dependence of the latter in
 the inset. Black centers identify an optimal rank $K=2$
 in the most probable block. Arrows indicate increasing mass.
 Dotted vertical lines in the mass plots give $m/g=0.3335$
 as a reference near the critical mass. }
 \label{fig:qis-massscan}
\end{figure*}

Varying the mass separates these changes from uncertainty
between vacua with opposite mean electric fields. 
Figure~\ref{fig:qis-massscan} compares half-ring entanglement
at $\theta=0$ and $\pi$ for $gL=20$ and $40$.
The total entropy and the within-sector entropy peak
near the critical mass for $\theta=\pi$, where the results are finite volume limited (due to the diverging correlation length). The decomposition as a function
of $\theta$, the sector weights and the leading entanglement
spectrum are shown in Appendix~\ref{app:qis-vacuum-anatomy} for completeness. The entanglement entropy in the $\theta$-vacuum was also investigated in~\cite{Grieninger:2026pai}.

Two additional spectral quantities in
Fig.~\ref{fig:qis-massscan} distinguish different aspects
of nonstabilizerness. The antiflatness is
\begin{equation}
 \mathcal A=\sum_k w_k
 \left[\operatorname{tr}\sigma_k^3
       -\bigl(\operatorname{tr}\sigma_k^2\bigr)^2\right].
 \label{eq:afdef}
\end{equation}
Inside each block, the antiflatness is the variance of
$\lambda_{ki}$ when $i$ has probability $\lambda_{ki}$.
It vanishes for a flat spectrum. A positive value witnesses a nonflat flux-resolved spectrum and
hence nonlocal magic within at least one flux sector%
~\cite{Tirrito:2023fnw}. Its magnitude weights spectral deviations
differently from the canonical value
$\mathcal M_2=\sum_k w_k\mathcal M_{2,k}$, so their extrema
need not coincide.

The nonlocal trace-distance magic compares a flux-resolved
pure state with the nearest stabilizer state after local
basis changes~\cite{Ebner:2026kwq}.
For a flat Schmidt spectrum of rank $K$, aligning the
Schmidt vectors gives the largest squared overlap
$F_{K,k}$. Pure-state trace distance then gives
\begin{equation}
 \begin{split}
 F_{K,k}&=\frac1K\left(\sum_{i=1}^{K}\sqrt{\lambda_{ki}}\right)^2,
 \qquad K=1,2,4,\ldots,\\
 \mathcal M_{\rm dist}^{\rm NL}
 &=\sum_k w_k\sqrt{1-\max_K F_{K,k}} .
 \end{split}
 \label{eq:mdistdef}
\end{equation}
The antiflatness bounds nonlocal magic from below,
$\mathcal A/8\leq\mathcal M_{\rm dist}^{\rm NL}$~\cite{Cao:2024nrx}.
It resolves spectral structure that $\mathcal M_2$ alone
does not determine. In particular, changes in the optimal
rank produce cusps and the maxima of the two quantities
need not occur at the same mass. 
Fig~\ref{fig:qis-massscan} shows how the entanglement entropy and magic measures change as a function of $m/g$ for $\theta=0$ (blue) and $\theta=\pi$ (red). The red curves increase with mass and show a peak of all measures near the critical point. 
In particular, $\mathcal M_{\rm dist}^{\rm NL}$ seems to resolved the critical point best at our given resolution as shown in panel (d) of Fig.~\ref{fig:qis-massscan}. As shown by panel (c) of Fig.~\ref{fig:qis-massscan}, magic is mostly tied to entanglement except for the the critical region where the dependency changes its scaling law.

\subsection{A common capacity minimum in the string and vacuum}
\label{sec:capacity-pair-production}

The capacity of entanglement measures the spread of the
modular energies, the eigenvalues of $K_A=-\ln\rho_A$~\cite{DeBoer:2018kvc}.
For a pure bipartite state, a nonzero capacity witnesses
nonlocal magic~\cite{Cao:2024nrx}.
Within block $k$, a modular energy is
$-\ln w_k-\ln\lambda_{ki}$. The corresponding flux-resolved mean is
$S_k-\ln w_k$ and the flux-resolved variance is the block
capacity $C_k$. The law of total variance therefore gives
\begin{align}
 C_A&=\operatorname{tr}(\rho_AK_A^2)-S_A^2
 \nonumber\\
 &=\sum_k w_k C_k+
   \sum_k w_k\bigl(S_k-\ln w_k-S_A\bigr)^2,
 \label{eq:qis-capacity}\\
 C_k&=\operatorname{tr}\!\left[\sigma_k(\ln\sigma_k)^2\right]-S_k^2 .
 \nonumber
\end{align}
The first term describes fluctuations within sectors.
The second describes differences between their mean
modular energies. A flat nonzero spectrum has zero capacity.

\begin{figure}[t]
 \centering
 \includegraphics[width=\columnwidth]{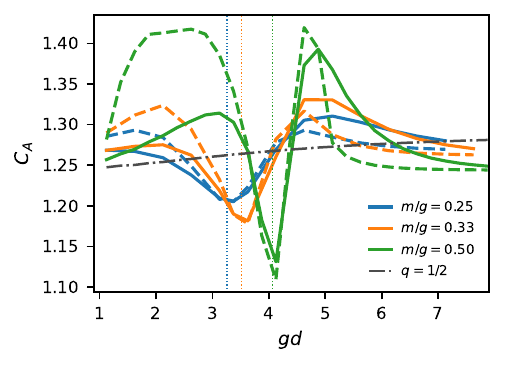}
 \caption{Capacity of the four-site central string interval
 (solid) and a four-site vacuum interval (dashed), for
 $m/g=0.25$, $0.33$ and $0.5$.
 The vacuum capacity is evaluated at the mean-flux-matched
 angle $\theff(d)$ of Eq.~\eqref{eq:qis-theta-match}.
 Dotted lines correspond to $\dbreak$.
 The gray dash-dotted curve is the $q=1/2$ string
 at $m/g=0.5$.
 Both use $ga=1/8$. }
 \label{fig:string_vs_vacuum}
\end{figure}

At $m/g=0.25$, $0.33$ and $0.5$, the central capacity falls
below its $\theta=0$ vacuum value near $\dbreak$.
Throughout this dip, the weighted within-sector capacity
exceeds its vacuum value. The decrease thus comes from
a smaller dispersion of the mean modular energies between sectors.
The massless string has only a small, smooth change from
its vacuum value and the $q=1/2$ string has no dip at all (see Fig.~\ref{fig:string_vs_vacuum}).

The binary contribution is simplest for a density matrix
consisting of two orthogonal blocks with probabilities
$p$ and $1-p$ and identical normalized spectra.
If that common spectrum has entropy $S_0$ and capacity
$C_0$, Eq.~\eqref{eq:qis-capacity} reduces to
\begin{align}
 S_A&=S_0+h_2(p),
 \nonumber\\
 C_A&=C_0+p(1-p)\left[\ln\frac{p}{1-p}\right]^2.
 \label{eq:qis-identical-blocks}
\end{align}
At equal weight the binary entropy $h_2$ is maximal and its
capacity vanishes. An exact Bell factor therefore adds $\ln2$ to
the residual entropy of the state while leaving its capacity unchanged. The vacuum at $\theta=\pi$ has the analogous approximate Bell
structure (Sec.~\ref{sec:vacuum-branch-qis}). The depth of a total
capacity minimum also depends on the flux-resolved spectra and the
remaining flux sectors.

The position $gd$ of the full capacity minimum depends on
both the sector probabilities and their internal spectra.
For two blocks with different entropies, their mean modular
energies agree when
$\ln[p/(1-p)]=S_I-S_B$.
The remaining sectors and the variation of $C_k$ further
affect the minimum. In the present scan, the lowest
capacities lie just beyond $\dbreak$, but the separation
step $\Delta(gd)=0.25$ does not resolve a common positive
displacement. We find that the capacity minimum occurs precisely near $\dbreak$ unresolved.

To compare with the uniform vacuum, we use an interval of
the same four sites and the same spacing on a ring with
$N=320$. We match the mean dynamical flux at the boundaries.
Writing $L_\partial=(L_L+L_R)/2$, the effective angle is
defined by
\begin{equation}
 \overline{\langle L_\partial\rangle}_{\rm vac}
       \bigl(\theff(d)\bigr)
 =\overline{\langle L_\partial\rangle}_{\rm string}(d).
 \label{eq:qis-theta-match}
\end{equation} 

Figure~\ref{fig:string_vs_vacuum} shows capacity minima of similar
shape and depth near flux exchange in both settings at all three masses.
Near string breaking, the matched vacuum angle is close
to $\pi$. Matching the mean flux does not constrain the
modular-energy variance, so the similar capacity
minima provide an independent spectral connection between
local string discharge and global vacuum flux exchange. At $m/g=0.5$, resolving
the narrow finite-ring vacuum transition permits this
comparison through the exchange. The curves differ
appreciably on the intact side, so the correspondence
is strongest near pair-induced discharge. It concerns
the spectra of the matched intervals at the stated
sizes and spacing.

In both settings, pair creation and charge separation change
the flux by one unit. The pair screens the external sources in
the string; charge transport around the ring connects neighboring
vacuum flux branches (Sec.~\ref{sec:model}). We interpret the
similar capacity minima as an entanglement signature of this shared
pair-creation mechanism. Their shape and depth probe the accompanying
spectral rearrangement beyond the binary entropy of equal branch
probabilities.

\section{Spatial correlations and nonlocal magic}
\label{sec:spatial}

The screening bit is shared by a whole interval and its complement.
Two small matter regions can retain different correlations because
the discarded degrees of freedom may still distinguish the flux
branches, removing their interference from the two-region state.
We examine these separated regions to determine how charge
rearrangement extends along the string%
~\cite{Grieninger:2026bdq}. We distinguish coherence between
different charge distributions from nonlocal magic. The first accounts
for most of the extra mutual information near breaking. The second
asks whether those correlations require a nonstabilizer resource
after separate changes of basis in the two regions.

\subsection{Charge correlations along the string}
\label{sec:charge-coherence}

\begin{figure*}[tbp]
\centering
\includegraphics[width=\textwidth]{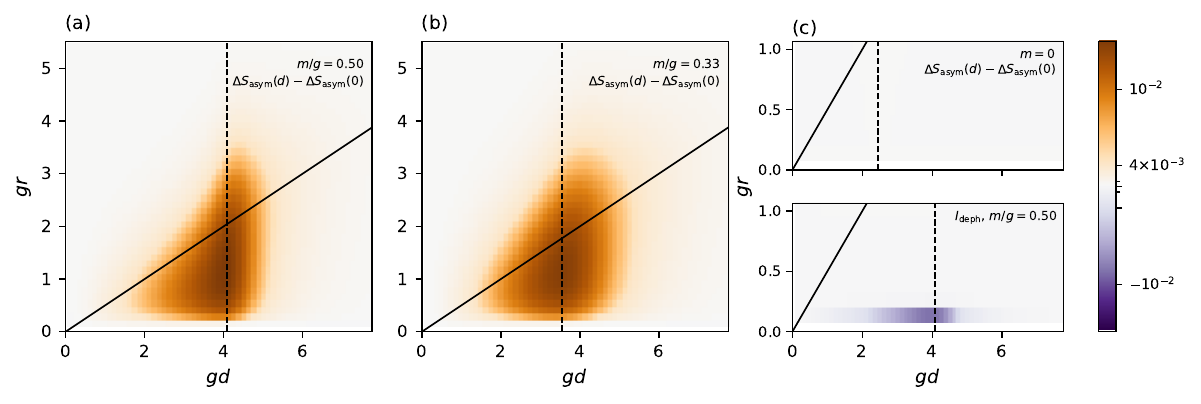}
\caption{(a), (b) The vacuum-subtracted
charge coherence $\DSasym(d,r)-\DSasym(0,r)$ at $m/g=0.50$ and
$0.33$ for the string using $N=320$ and $ga=1/16$. Region $A$ is at the string
midpoint and $B$ is displaced by $r$. (c) The same quantity at $m=0$ (upper panel) and the
vacuum-subtracted $I_{\rm deph}$ at $m/g=0.50$ (lower panel).
Solid lines give $r=d/2$, where $B$ reaches an external charge.
The massive dashed lines mark $\dbreak$.}
\label{fig:spatial-fine}
\end{figure*}

\begin{figure*}[tbp]
\centering
\includegraphics[width=\textwidth]{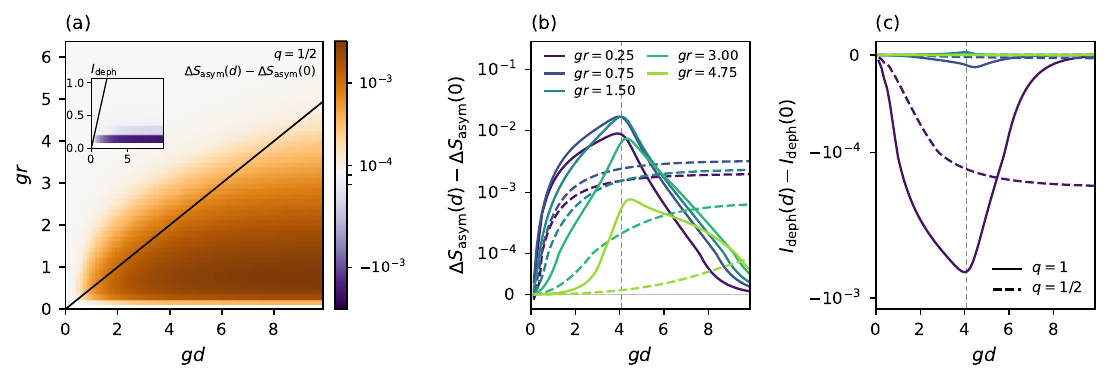}
\caption{ (a) The difference
$\DSasym(d,r)-\DSasym(0,r)$ for the $q=1/2$ string in the same setting as Fig.~\ref{fig:spatial-fine}. The corresponding
$I_{\rm deph}(d,r)-I_{\rm deph}(0,r)$ in the inset.
(b),(c) Comparison of spatial slices for $q=1$ and $q=1/2$ at the same mass.}
\label{fig:spatial-halfcharge}
\end{figure*}

We follow the setup in~\cite{Grieninger:2026bdq} and place a small region $A$ at the string midpoint and a second
region $B$ a distance $r$ to its right. The distance between the
external charges is $d$, so $B$ reaches the right-hand charge
at $r=d/2$. Each region contains one physical site i.e. two staggered
sites, giving four occupation states per region. We use the
spin reduced density matrix $\rho_{AB}$, obtained after solving
Gauss's law and expressing the fermions through the Jordan--Wigner
map. For separated regions, this differs from the
fermionic reduction~\cite{Igli2009OnRD}. They differ in
off-diagonal elements with odd changes of local charge.
For a fermion correlation joining the inner edges of the two
regions, the reduced density matrix sums the contributions from
even and odd occupations with a sign depending on the occupation of the gap. The spin reduced density matrix is agnostic to the gap parity. Gauss's law relates this occupation parity
to the difference of electric fields at the gap boundaries. The spin correlation
therefore tests how coherence between the regions depends on
the charge and electric field in the gap. Both correlations
include the Wilson line required by gauge invariance.
Appendix~\ref{app:spatial} includes their precise definitions and a comparison.

The mutual information,
$I(A{:}B)=S(\rho_A)+S(\rho_B)-S(\rho_{AB})$, includes both
correlations of the local charges and coherence between different
charge distributions. To distinguish these contributions, we remove
matrix elements connecting different values of $Q_A$ or $Q_B$:
\begin{equation}
 \mathcal D(\rho_{AB})=
 \sum_{q_A,q_B}P_{q_A,q_B}\rho_{AB}P_{q_A,q_B}.
 \label{eq:spatial-dephasing}
\end{equation}
Here $P_{q_A,q_B}=P^A_{q_A}\otimes P^B_{q_B}$, where
$P^A_{q_A}$ projects onto charge $q_A$ in $A$ and similarly
for $B$. This dephasing preserves the charge probabilities and
the density matrix within each fixed-charge block.
Charge conservation already makes $\rho_A$ and $\rho_B$ block
diagonal in their respective charges. Their entropies are therefore
unchanged and the loss of mutual information equals the increase
of the joint entropy:
\begin{align}
 \DSasym&=S(\mathcal D\rho_{AB})-S(\rho_{AB}),
 \label{eq:spatial-asymmetry}\\
 I(A{:}B)&=I_{\rm deph}(A{:}B)+\DSasym.
 \label{eq:spatial-mi-split}
\end{align}
The joint state can retain phase coherence between different
charge distributions even when it is mixed. Dephasing removes
this coherence and leaves a statistical mixture of regional
charges with the same probabilities. Thus $\DSasym$ is the part of
the mutual information that depends on coherence between
different regional charges. Gauss's law fixes the change of
electric field accompanying each charge rearrangement.
The quantity equals the nonnegative
relative entropy $D(\rho_{AB}\Vert\mathcal D\rho_{AB})$
\cite{Gour:2009hng}. The remaining $I_{\rm deph}$
includes correlations of the charge probabilities as well as
quantum correlations within fixed-charge blocks.

To compare the string with the vacuum, we subtract the $d=0$
value, i.e.
$\DSasym(d,r)-\DSasym(0,r)$. We make the same subtraction
for $I$ and $I_{\rm deph}$. Positive and negative differences
denote enhancement and suppression of the corresponding quantity
relative to the vacuum. In particular, a negative charge-coherence
difference means less coherence than in the vacuum.

At $m/g=0.33$ and $0.50$, the enhanced mutual information near
breaking connects separated regions across much of the string
(Fig.~\ref{fig:spatial-fine}). Almost all of this increase is
accounted for by $\DSasym$, while the vacuum-subtracted
$I_{\rm deph}$ is small. Removing the coherence between regional
charges therefore removes most of the excess information shared
by the regions, while preserving their charge probabilities.
The enhancement detects a change in the quantum state of the
charges and their electric field that charge probabilities alone
do not describe. 

In the massless case, the charge coherence remains close to its vacuum
value throughout the plotted range (see Fig.~\ref{fig:spatial-fine}(c)). The electric field is screened without the extended enhancement of charge coherence seen in the massive strings.

The decrease after breaking follows the change from a flux tube
to two screened external charges. As the screening clouds separate,
a fixed region in the interior recovers its vacuum correlations.
The half-charge string tests the role of this discharge. At
$m/g=0.50$, an integer-charge pair cannot neutralize external
charges of magnitude $g/2$ and a residual string remains.
In Fig.~\ref{fig:spatial-halfcharge}, its coherence between
separated probes grows toward a sustained value as $d$ increases.
For unit charges, the corresponding curves reach a maximum and
then fall. Increasing the separation of the charges therefore
does not by itself explain the loss of excess coherence. The
decrease accompanies screening of the interior after the string has broken,
while the residual half-charge string continues to support
correlations across that interior.

\subsection{Spatial correlations in the $\theta$-vacuum}

\begin{figure*}[tbp]
\centering
\includegraphics[width=\textwidth]{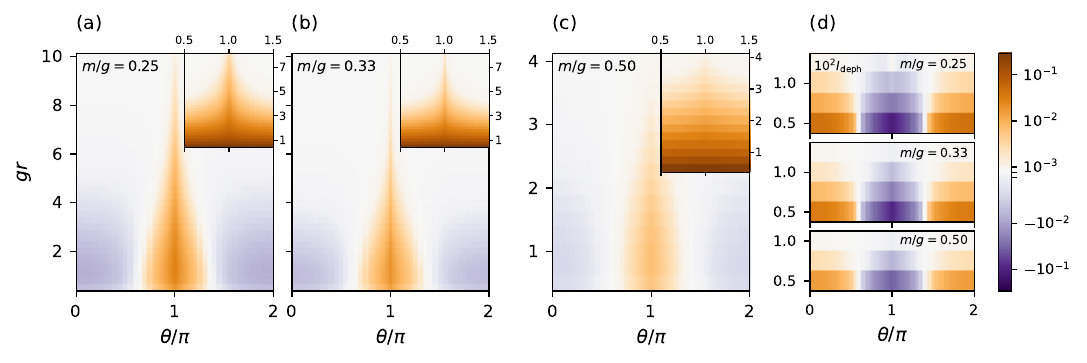}
\caption{Charge coherence in the $\theta$-vacuum of two regions of 1 physical site each at $N=320$ and $ga=1/8$. 
(a)--(c) $\DSasym(\theta,r)$ after subtracting its average
over $\theta$ at each fixed $r$. Insets show the unsubtracted
$\DSasym$ near $\theta=\pi$. (d) The corresponding change of
$I_{\rm deph}$ relative to its average over $\theta$, multiplied
by $10^2$.}
\label{fig:spatial-vacuum}
\end{figure*}

In the periodic vacuum, the same construction
shows how flux exchange reorganizes charge
coherence without external charges or their screening clouds.
As $\theta$ approaches $\pi$, $\DSasym$ increases between
separated regions (see Fig.~\ref{fig:spatial-vacuum}). This response
extends to distances where the change of $I_{\rm deph}$ is
already small. As in the string, most of the enhancement would be lost by
replacing the state with a mixture of definite local charges.
Coherent charge rearrangement thus accompanies both finite-size breaking of the string and the global vacuum flux exchange. In the presentation of the figure we chose to subtract the mean of the quantity over $\theta$ at a fixed interval separation $gr$ due to large differences in correlation length across $\theta$. This differs from the string setup where the $d=0$ vacuum is an appropriate reference.

Global branch uncertainty and this spatial charge coherence have
different mass dependence. Among the three masses shown,
$\DSasym$ near $\theta=\pi$ is strongest and persists to the
largest probe separations at $m/g=0.25$, even though the vacuum
branch exchange becomes sharper as the mass increases.

\subsection{Nonlocal magic in the string and vacuum}
\label{sec:spatial-magic}

Entangled stabilizer states can retain coherence between
different regional charges.
Nonstabilizerness tests a further property of the state: whether
it can be represented as a probabilistic mixture of stabilizer
states in the chosen qubit encoding. For the mixed state of the
two regions, we use the robustness of magic (RoM)
\cite{ Howard:2017maw,Pashayan:2015cos},
\begin{equation}
 \mathfrak R(\rho)=\min_{\{c_j\}}
 \left\{\sum_j|c_j|:\ \rho=\sum_jc_j\tau_j\right\},
 \label{eq:spatial-robustness}
\end{equation}
where $\tau_j$ range over all pure four-qubit Pauli stabilizer
density matrices and the real
coefficients sum to one. A probabilistic mixture has
$\mathfrak R=1$. Negative coefficients are necessary for a
nonstabilizer state, giving $\mathfrak R>1$.
We denote the excess in the original occupation basis by
$R_{\rm plain}=\mathfrak R-1$.

Nonlocal magic is the part that survives separate basis changes
within the two regions~\cite{Qian:2025oit}. Following
Ref.~\cite{Grieninger:2026bdq}, we restrict these changes
to preserve the local charges:
\begin{equation}
 R_{\rm NL}(\rho_{AB})=
 \min_{U_A,U_B\in\mathcal G_Q}
 \mathfrak R\!\left(U\rho_{AB}U^\dagger\right)-1,
 \quad U=U_A\otimes U_B.
 \label{eq:spatial-nl-robustness}
\end{equation}
Here $\mathcal G_Q$ contains all unitaries commuting with the
charge of the corresponding region. These transformations commute
with $\mathcal D$ and leave $\DSasym$ unchanged. The minimization
therefore tests nonstabilizerness without changing the amount of
charge coherence. For an uncorrelated state
$\rho_A\otimes\rho_B$, an allowed transformation diagonalizes
each charge-diagonal factor in the occupation basis. The result
is a probabilistic mixture of computational-basis states and
has $R_{\rm NL}=0$. A nonzero minimum therefore identifies
nonstabilizerness shared by the two regions.
The figures report the smallest values found in the basis search,
denoted $\widehat R_{\rm NL}$ below. 

The choice of reduced density matrix matters for this resource measure. For the fermionic reduced density matrix of the $m/g=0.50$ string states, we find zero $\widehat R_{\rm NL}$. The spin reduced density matrix contains a finite $\widehat R_{\rm NL}$ as described in the following. 

\begin{figure}[tbp]
\centering
\includegraphics[width=\linewidth]{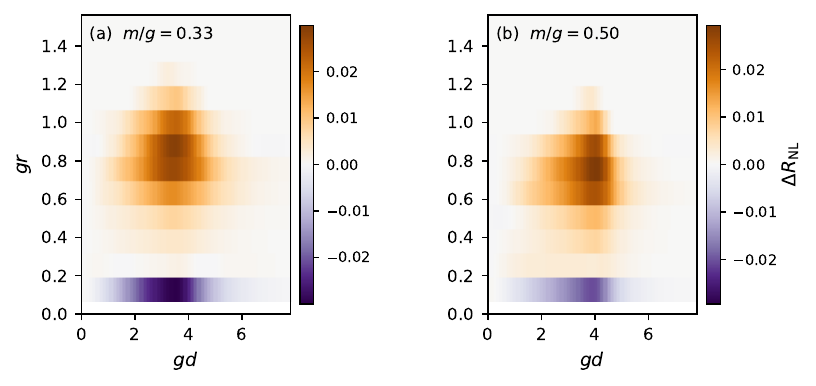}
\caption{Vacuum-subtracted nonlocal RoM of the strin,
$\Delta R_{\rm NL}(d,r)=\widehat R_{\rm NL}(d,r)
-\widehat R_{\rm NL}(0,r)$, at $\theta=0$, $N=320$ and
$ga=1/16$. We keep the reference interval $A$ corresponding to two physical sites centered between the
external charges and the other interval B is a distance $r$ to its right.
The three panels show $m/g=0$, $0.33$ and $0.50$.}
\label{fig:spatial-string-nl-fine}
\end{figure}

For the string, Fig.~\ref{fig:spatial-string-nl-fine} shows the
difference from the vacuum,
$\Delta R_{\rm NL}(d,r)=\widehat R_{\rm NL}(d,r)
-\widehat R_{\rm NL}(0,r)$.
At $m/g=0.33$ and $0.50$, there is a clear enhancement near breaking. 
The enhancement is concentrated closer to the midpoint than the charge-coherence signal in Fig.~\ref{fig:spatial-fine}. Its maximum moves to larger $d$ with increasing mass and generally precedes the equal-weight $d_\star$. The largest excess therefore occurs in states on the approach to breaking. At larger $d$, the probes lie between separated screening clouds and recover the correlations of the vacuum. The results match our previous findings in~\cite{Grieninger:2026bdq}

In the homogeneous vacuum, varying the separation of the regions
leaves their individual reduced states unchanged. The dependence
of $R_{\rm plain}$ on $r$ therefore comes from correlations
between them. Each region can also retain magic on its own. We remove this floor by subtracting the value at a large reference separation:
\begin{equation}
 c_\theta(r)=R_{\rm plain}(\theta,r)
             -R_{\rm plain}(\theta,r_{\rm ref}).
 \label{eq:spatial-plain-subtraction}
\end{equation}

\begin{figure*}[tbp]
\centering
\includegraphics[width=\textwidth]{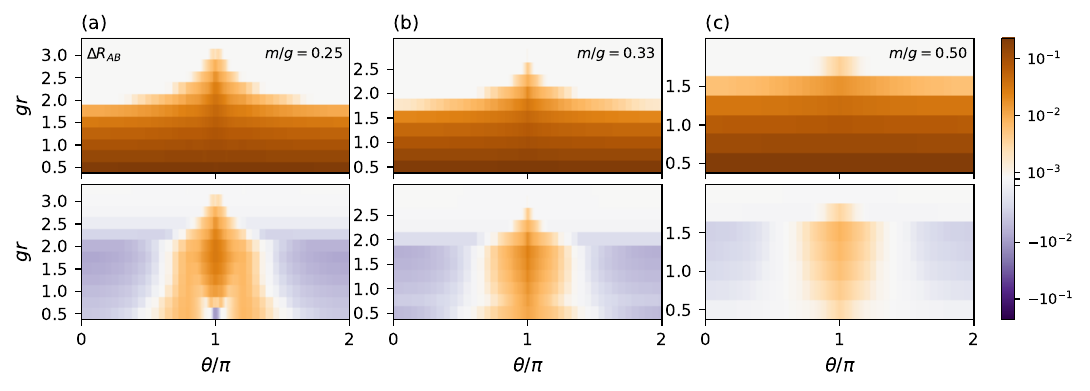}
\caption{ RoM of the $\theta$-vacuum for $N=320$ and $ga=1/8$.
We subtract a reference value at large separation $r_{\rm ref}$,
$c_\theta(r)=R_{\rm plain}(\theta,r)
-R_{\rm plain}(\theta,r_{\rm ref})$ to remove the RoM of the individual regions.
Upper panels are the raw values. Lower panels subtract
the average over $\theta$ at each fixed $r$. The columns show $m/g=0.25$, $0.33$ and $0.50$. }
\label{fig:spatial-vacuum-plain}
\end{figure*}

Figure~\ref{fig:spatial-vacuum-plain} shows that $R_{\rm plain}$ is the longest range near $\theta=\pi$ modify $R_{\rm plain}$. 

\begin{figure*}[tbp]
\centering
\includegraphics[width=\textwidth]{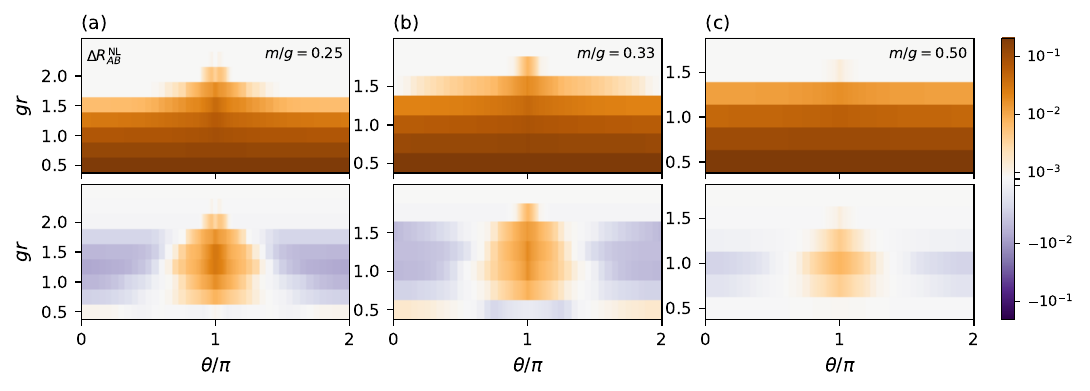}
\caption{Nonlocal RoM of the $\theta$-vacuum for $N=320$ and $ga=1/8$. Upper panels show the raw $\widehat R_{\rm NL}(\theta,r)$.
Lower panels subtract the average over $\theta$ at each fixed $r$. The columns show $m/g=0.25$,
$0.33$ and $0.50$.  }
\label{fig:spatial-vacuum-nl}
\end{figure*}

The enhancement near $\theta=\pi$ persists when computing the nonlocal RoM by minimization and is shown in  Fig.~\ref{fig:spatial-vacuum-nl}. 

\section{Neural representation of vacuum and string states}
\label{sec:neural-states}
In this section, we construct the wave functions of string and vacuum
with a restricted Boltzmann machine (RBM) which is a neural quantum state
with one layer of hidden binary variables~\cite{Carleo:2016svm}. The section focuses on the architecture and construction of the RBMs. While tensor networks work particularly well in one dimension due to the favorable scaling of entanglement, the extension to two spatial dimensions is resource intensive. Neural quantum states may not be restricted by the entanglement growth. We hence explore their construction in the current setup with extensions to higher dimensions in mind. Our construction combines a shared learned embedding of the electric fluxes reconstructed through Gauss's law with translation sharing on the ring. Releasing the sharing of the network weights extends
the vacuum ansatz to the inhomogeneous string. We optimize the energy and compare the resulting states with exact diagonalization (ED). These calculations test the representation at finite lattice size, where the wavefunction can be evaluated on the entire physical basis.

RBMs with binary spin inputs have been applied to the Schwinger model~\cite{Crew:2025khs}. Other gauge-theory neural states use embedded matter--flux configurations with explicit Gauss-law checks~\cite{Luo:2021gls}, or gauge-equivariant networks on group-valued links~\cite{Luo:2020stn}. Here the physical basis already solves Gauss's law. Reconstructing the fluxes supplies features adapted to spatial translations without restoring independent gauge variables.

\subsection{An RBM with reconstructed flux inputs}
\label{sec:rbm-architecture}

In the neutral sector, a basis configuration consists of the fermion occupations $\boldsymbol n$ and one integer reference flux $L_{\rm ref}$. Gauss's law reconstructs the remaining fluxes from these independent variables. For the network we relabel the sites from zero, so site $i$ is site $n=i+1$ of Eq.~\eqref{eq:H}. At staggered parity zero the
reconstructed inputs are
\begin{equation}
 L_n=L_{\rm ref}+\sum_{i=0}^{n}(n_i-v_i),\qquad
 v_i=\frac{1+(-1)^i}{2}.
 \label{eq:neural-link-input}
\end{equation}
Here $\sum_i n_i=N/2$, so the last link returns to $L_{\rm ref}$. The sequence retains the full basis configuration: its last entry gives $L_{\rm ref}$ and neighboring differences recover the charges. The interacting vacuum superposes neutral configurations with charge fluctuations, so its individual configurations need not have uniform $L_n$. The prescribed background and source fields enter the total electric field in the Hamiltonian, Eq.~\eqref{eq:fielddef}, rather than the embedding indices.

A change of $L_{\rm ref}$ shifts every reconstructed flux, while a charge pair changes the flux over the interval between its charges. Translations act by cyclically shifting this sequence, even though they can change $L_{\rm ref}$ in the original coordinates. These inputs therefore treat the reference link consistently with the rest of the ring and make the spatial translation symmetry of the vacuum simple to impose.

Each integer $L_n$ is mapped to a learned complex vector $\boldsymbol e(L_n)$ with four components. For network
parameters $p$, the unnormalized amplitude is
\begin{align}
 \psi_p(\boldsymbol n,L_{\rm ref})={}&
 \exp\!\left[\sum_n\boldsymbol b^{\rm v}_n\cdot
                         \boldsymbol e(L_n)\right]
 \nonumber\\[-2pt]
 &\times\prod_{h=1}^{N_h}2\cosh\!\left[
 b^{\rm h}_h+\sum_n\boldsymbol W_{hn}\cdot\boldsymbol e(L_n)
 \right].
 \label{eq:neural-rbm}
\end{align}
The visible biases $\boldsymbol b^{\rm v}_n$ weight the embeddings at individual links. Each hidden variable couples to the reconstructed flux configuration through $\boldsymbol W_{hn}$. Summing that variable over $\pm1$ gives its $2\cosh$ factor. The embedding, weights and biases are complex, allowing the network to describe the phase as well as the magnitude of the wavefunction. The embedding is a lookup table over the allowed integer inputs.

The vacuum is invariant under translations by two staggered sites. We impose this by translating each of eight independent coupling patterns through the $N/2$ cells and multiplying their hidden-unit factors. The visible biases repeat every two links. This gives $N_h=4N$ hidden units while keeping the two links within a cell inequivalent. The number of independent parameters grows linearly with $N$ at fixed embedding dimension and reference-flux cutoff. Appendix~\ref{app:neural-details} gives the sharing relations and parameter counts. Only $L_{\rm ref}$ is cut off independently;
the remaining fluxes follow from Gauss's law.

\subsection{Vacuum optimization and the low-energy pair}
\label{sec:rbm-vacuum}

 For each mass, the vacuum scan starts at
$\theta=0$ and uses trained states at nearby angles to initialize
the remaining points. We retain the lowest-energy recorded
parameters; ED supplies an independent test of the resulting state.

\begin{figure}[tbp]
 \centering
 \includegraphics[width=\columnwidth]{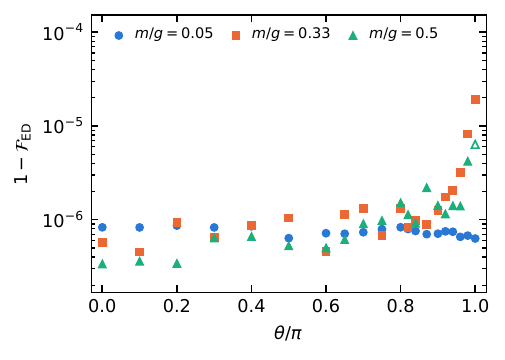}
 \caption{Vacuum infidelity of the RBM for $N=24$,
 $ga=1$, $L_{\max}=5$ and the three masses indicated. Adam
 optimization is followed by L-BFGS refinement. The reference wavefunctions were obtained by ED.
}
 \label{fig:neural-vacuum-fidelity}
\end{figure}

For a normalized neural state $|\Psi\rangle$, the squared overlap
with the normalized ED ground state,
\begin{equation}
 \mathcal F_{\rm ED}=|\langle\psi_{\rm ED}|\Psi\rangle|^2,
 \label{eq:neural-fidelity}
\end{equation}
tests the full wavefunction. At $N=24$ and $ga=1$, the
optimized single RBM has $1-\mathcal F_{\rm ED}<2\times10^{-5}$
over the considered angles (Fig.~\ref{fig:neural-vacuum-fidelity}). Over this
range, the reference-flux probabilities also agree with ED, testing the
global flux distribution represented by the network.

\subsection{The string RBM}
\label{sec:string-rbm}

External charges break translation invariance and produce
different field distributions along the ring. We allow the
visible biases, hidden biases and coupling patterns in
Eq.~\eqref{eq:neural-rbm} to vary independently with position, retaining the shared embedding and the same hidden-unit count. This permits the amplitude to distinguish the string interior, the screening clouds and the exterior. Copying the translated
vacuum parameters into the independent weights initially preserves its wavefunction exactly. Subsequent energy minimization
can then deform it around the charges.

For $N=24$, $ga=0.5$ and $m/g=0.33$, we first optimize the
vacuum and use it to initialize the shortest unit string,
$d/a=1$. All larger separations start independently from this
optimized short-string state, with fresh optimizer variables.
The two charges remain centered about the same midpoint. Each string is represented by a single RBM. Adam optimization
of the full-sum energy is followed by L-BFGS refinement and we compare the lowest-energy parameters recorded across both stages
with a separately computed ED ground state.

\begin{figure}[tbp]
 \centering
 \includegraphics[width=\columnwidth]{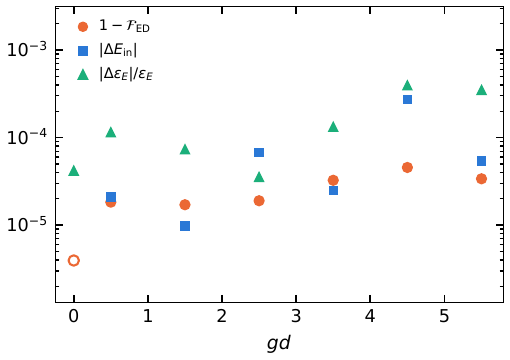}
 \caption{String infidelity and expectation value errors
 at the lowest-energy parameters,
 for $N=24$, $ga=0.5$, $m/g=0.33$, $L_{\max}=5$, $q=1$ and
 $\theta=0$. Circles show $1-\mathcal F_{\rm ED}$;
 squares show the absolute $\Ein$ error
 $|\Ein^{\rm RBM}-\Ein^{\rm ED}|$. Triangles show the relative error
 $|\varepsilon_E^{\rm RBM}-\varepsilon_E^{\rm ED}|/
 \varepsilon_E^{\rm ED}$, where
 $\varepsilon_E=N^{-1}\sum_n\langle E_n^2\rangle$ is the
 electric energy density in units of $g^2/2$.
 ED supplies the reference wavefunctions.}
 \label{fig:string-rbm-fidelity}
\end{figure}

The optimized states reproduce both the string wavefunction
and its electric observables across the considered separations
(Fig.~\ref{fig:string-rbm-fidelity}). The infidelity remains
below $5\times10^{-5}$; the midpoint field and electric energy
density agree with ED to within $0.15\%$ and $0.04\%$,
respectively. Releasing translation sharing thus accommodates
the inhomogeneous field and its screening with the same
reconstructed flux inputs and hidden-unit density used for
the vacuum.

\section{Conclusions and outlook}
\label{sec:conclusions}

Pair production links the discharge of a string to the rearrangement of global vacuum flux. In the Schwinger model, a dynamical pair changes the field by one unit between its charges. The screening charges are localized to the external charges in case of the string, while charge transport around the ring
connects neighboring vacuum branches. Studying both setups across
the fermion mass range shows how this common process changes from
smooth polarization to competition between sharply distinguished
flux configurations. The electric field, excitation spectrum and
spatial entanglement describe complementary parts of that mechanism.

The midpoint electric field gives an order parameter to characterize string breaking similarly to its role for the $\theta$ vacuum.
We define the separation $\dbreak$ of the external charges where the string breaks where the midpoint electric field has fallen to its half-unit. Since the string is a finite size object, string breaking generally extends over a range of separations characterized by the slope of the midpoint electric field at breaking and $\dbreak$ is understood as the midpoint of the transition. This is analogous to $\theta=\pi$ below $m_c$ where the vacuum exhibits a crossover.  At finite mass, the crossing of modeled intact and
screened energies reproduces $\dbreak$ using only independent vacuum and static-light inputs. Screening clouds contribute to both energies and remain essential when their extent is comparable to the separation. By computing the flux-resolved weights at the midpoint, we find that $\dbreak$ also coincides with equal probability of the intact string and broken, screened configuration. This equal probability results in a maximal entanglement entropy of the binary flux label. Hence, flux uncertainty accounts for most of the entanglement entropy at string breaking and the vacuum transition.

The excited states determine how sharply the flux changes with separation. At light mass, the lowest excited state is a meson bound by the screening profile. At larger mass, the two lowest states exchange their interior fields through an avoided crossing and the location of the minimum gap approaches $\dbreak$.

An external electric field anti-aligned with the string makes the connection to the $\theta$ vacuum explicit. Above the critical mass, the string $\dbreak$ diverges for $\theta\to-\pi$ and the mass gap closes. The string smoothly turns into the $\theta$ vacuum since it also looses its tension in this limit.
Below the critical mass, the unique vacuum instead supports a massive meson that can become the lowest excitation by occupying the exterior.

Screening organizes part of the entanglement between the string
interior and exterior into a nearly maximally entangled bit. It
records whether the flux is screened and nearly separates from
the remaining correlations after local encoding. The flux-resolved
spectra establish this approximate Bell factor for coherently
selected flux sectors near breaking, with an
analogous mechanism for the $\theta$-vacuum transition. Orthogonal
flux records make the relative phase inaccessible on either side; equal probabilities give the binary entropy $\ln2$. This gives
the MEL mechanism a concrete binary realization%
~\cite{Kharzeev:2026jkq,Grieninger:2025rdi}, with spectral matching establishing the further separation of the bit from its
surroundings. 
Entanglement within the intact, screened and vacuum branches also grows toward the transition, revealing correlations beyond the shared screening information. For the $\theta$-vacuum we find that nonlocal nonstabilizerness peaks near the critical point and changes character in terms of its relation to the entanglement entropy.

The capacity of entanglement supplies a further correspondence between the two setups. Matching the mean dynamical flux reveals minima of similar shape and depth in the central string and vacuum intervals at all three masses. The unbreakable $q=1/2$ string lacks this dip. This spectral
rearrangement contains information beyond equal binary weights.
We interpret the correspondence as evidence that the shared
pair-creation mechanism reorganizes spatial entanglement similarly
near string breaking and vacuum flux exchange.

As argued in~\cite{Grieninger:2026bdq}, to visualize the rearrangment of degrees of freedom, we have to consider disjoint regions.
The entanglement asymmetry of disjoint regions reveals coherence between different charge distributions and their associated electric fields. Removing
this coherence preserves the charge probabilities, yet removes
most of the excess mutual information
near string breaking and vacuum flux exchange. This can be quantified further by the (non-local) Robustness of Magic.

Finally, we constructed the string and $\theta$-vacuum wavefunctions using Neural Network Quantum States.
Combining Gauss-law reconstruction with a shared complex embedding gives an RBM representation of periodic vacuum and string wavefunctions at high fidelity for all parameters considered.

In summary, we put forward a criterion for string breaking in terms of the midpoint electric field that is independent of excited states and equilibrium energies. 
Hence, we expect this characterization to hold in non-equilibrium settings where a clean observable to parametrize string breaking is still missing. Furthermore, the criterion should also extend to non-abelian theories.
Euclidean lattice correlation matrices have resolved string--meson mixing and the distinct thresholds from light- and strange-quark production%
~\cite{Bali:2005fu,Bulava:2019iut}. Chromoelectric profiles provide a complementary spatial observable~\cite{Baker:2024peg}.
Combining them would test whether lost flux and the spectral mixing identify the same configuration exchange.  

\section*{Acknowledgements}
We thank Dima Kharzeev, Eliana Marroquin, Martin Savage and Nikita Zemlevskiy for insightful discussions and collaboration on related work. 
 This material is based upon work supported by the U.S. Department of Energy, Office of Science, National Quantum Information Science Research Centers, Quantum Systems Accelerator (Award No. DE-SCL0000121). Additional support is acknowledged from the U.S. Department of Energy, Office of Science, Office of Nuclear Physics, Inqubator for Quantum Simulation (IQuS) under Award Number DOE
(NP) Award DE-SC0020970 (S.G.). This work was also supported by the U.S. Department of Energy, Office of Science, Office of Nuclear Physics, Grant No. DE-FG02-97ER-41014 (UW Nuclear Theory, S.G.). S.G. was supported in part by a Feodor Lynen Research fellowship of the Alexander von Humboldt foundation. 
We acknowledge the use of {\tt Codex} and {\tt Claude}.
This work was also supported, in part, by the Department of Physics and the College of Arts and Sciences at the University of Washington.
This work was enabled, in part, by the use of advanced computational, storage and networking infrastructure provided by the Hyak supercomputer system at the University of Washington.
This research used resources of the National Energy Research Scientific Computing Center (NERSC), a Department of Energy Office of Science User Facility using NERSC award NP-ERCAP0032083.
This research was done using services provided by the OSG Consortium \cite{osg07, osg09, https://doi.org/10.21231/906p-4d78, https://doi.org/10.21231/0kvz-ve57}, which is supported by the National Science Foundation awards \#2030508 and \#1836650.
We have made use of the ITensor library for tensor network computations.
\appendix
\section{Numerical formulation and extraction}
\label{app:numerics}

The calculations use the periodic Hamiltonian in Eq.~\eqref{eq:H}
in the sector of zero total dynamical charge. Solving Gauss's law
expresses each link flux in terms of the fermion occupations and
one reference flux, as written explicitly in
Eq.~\eqref{eq:neural-link-input}. The latter is an integer degree
of freedom conjugate to the ring holonomy. It must remain dynamical
to describe the vacuum flux exchange. Hopping across the reference
link changes this integer along with the occupations.

The reference register is restricted to
$-L_{\max}\leq L_{\rm ref}\leq L_{\max}$, with
$L_{\max}=5$ for the information and neural-state results.
The other integer fluxes $L_n$ include accumulated dynamical charge
and are not cut off independently. The prescribed background and
source fields enter the total electric field separately, as in
Eq.~\eqref{eq:fielddef}. 

Ground states and low excitations at the larger volumes are
obtained by density-matrix renormalization group calculations
with matrix product states, using ITensor~\cite{ITensor}.
Excited states are optimized with penalties for overlap
with the states already obtained.

\section{Flux-resolved information and spectral conventions}
\label{app:qis}

\subsection{Constructing the flux blocks}
\label{app:qis-assembly}

We store each sector $k$ using integer dynamical fluxes
$(\ell_L,\ell_R)$. Between the external charges the total fields are
$E_{L,R}=\ell_{L,R}+q+\theta/(2\pi)$, so a $q=1$ string at has intact flux one at $(0,0)$ and screened flux zero at $(-1,-1)$. For the vacuum, set $q=0$.
For each fixed flux pair, choose orthonormal gauge-compatible
bases $|a;k\rangle$ in
the interval and $|b;k\rangle$ in its complement.
The amplitudes of the pure state form a matrix
$\Psi^{(k)}_{ab}$ in each such sector. Tracing the
complement gives
\begin{equation}
 \bigl(\rho_A^{(k)}\bigr)_{aa'}
 =\sum_b\Psi^{(k)}_{ab}\Psi^{(k)*}_{a'b},\qquad
 w_k=\sum_{ab}|\Psi^{(k)}_{ab}|^2.
 \label{eq:qis-block-assembly}
\end{equation}
The blocks are positive and their traces sum to one.
Their normalized eigenvalues define $\sigma_k$ and
$\lambda_{ki}$ in Eq.~\eqref{eq:qis-blocks}. Thus
the flux-resolved spectra retain all compatible matter
configurations within the specified boundary fluxes.
In the MPS calculation,
the same sums are contractions over the complement.

Resolving only the enclosed charge merges blocks with
the same $Q=\ell_R-\ell_L$. With
$p_Q=\sum_{k:Q_k=Q}w_k$, the normalized charge block is
$\bigoplus_{k:Q_k=Q}(w_k/p_Q)\sigma_k$. Its entropy includes the uncertainty of $k$ at fixed $Q$,
denoted $H(k|Q)$, as well as the entropies of the flux blocks.
With $H(Q)=-\sum_Qp_Q\ln p_Q$, the two terms become
\begin{equation}
 \begin{aligned}
 H(Q)&=\Sconf-H(k|Q),\\
 \sum_Qp_Q S(\rho_{A|Q})&=\Ssec+H(k|Q).
 \end{aligned}
 \label{eq:qis-charge-grouping}
\end{equation}
Their sum remains $S_A$. This accounts for the
different divisions into probability and flux-resolved
entropy when only charge is resolved. The flux pair
used here also distinguishes neutral sectors through
which different electric flux passes.

\subsection{Vacuum entropy across the flux exchange}
\label{app:qis-vacuum-anatomy}

\begin{figure*}[t]
 \centering
 \includegraphics[width=\textwidth]{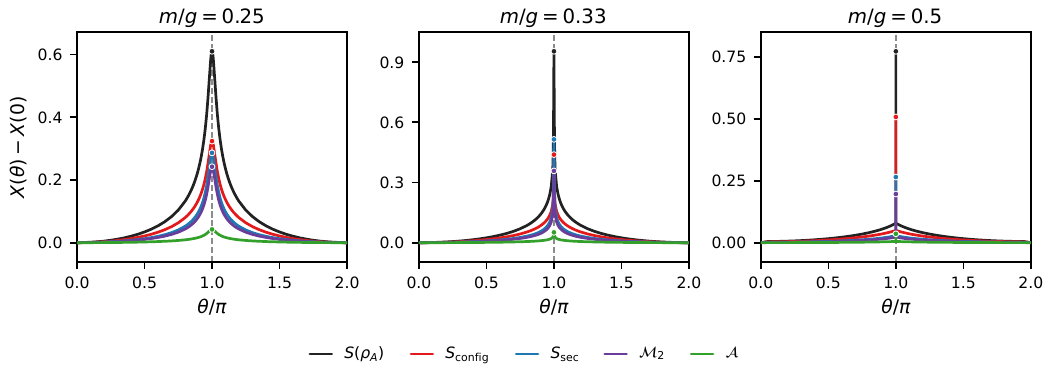}
 \caption{Vacuum quantum information as a function of $\theta$,
 relative to its value at $\theta=0$, for half-space with $N=320$, $ga=1/8$ and $L_{\max}=5$.
 The total entropy equals $\Sconf+\Ssec$.
 The other curves are the weighted flux-resolved
 $\mathcal M_2$ and antiflatness $\mathcal A$.
 Dots specify the values at $\pi$.}
 \label{fig:qis-vacuum-anatomy}
\end{figure*}

Figure~\ref{fig:qis-vacuum-anatomy} shows the half-space entanglement entropy.
Both the flux-probability entropy and the flux-resolved entropy grow near $\pi$. Their relative contributions
vary with mass. At $m/g=0.5$, much of the sharp increase
comes from the probabilities of the competing flux
orientations, while the flux-resolved entropy and
nonstabilizerness also increase.
This separates internal changes from the additional
uncertainty in the vacuum orientation.

\begin{figure*}[t]
 \centering
 \includegraphics[width=\textwidth]{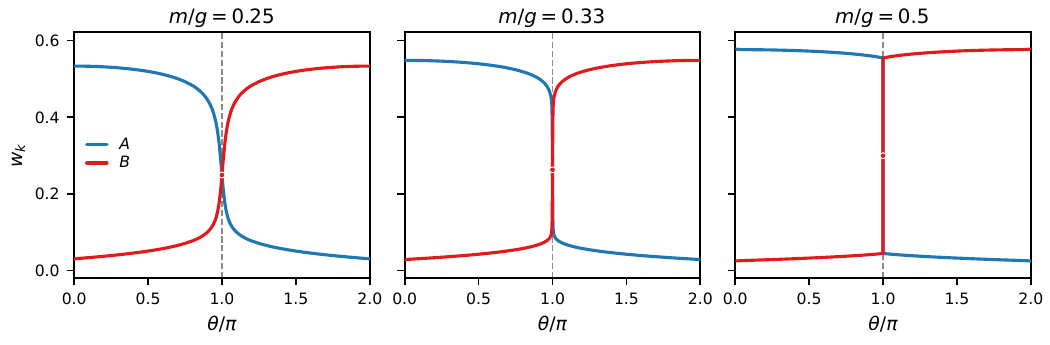}
 \caption{Absolute probabilities of the neutral vacuum branches
 $A$ and $B$, carrying total bounding flux $\theta/(2\pi)$ and
 $\theta/(2\pi)-1$, for the half-space intervals used in
 Fig.~\ref{fig:vac-branch-qis}.
 Dots give their weight at $\theta=\pi$.
 The curves use $N=320$, $ga=1/8$ and $L_{\max}=5$,
 with the same angular samples and interpolation as
 Fig.~\ref{fig:qis-vacuum-anatomy}.}
 \label{fig:qis-vacuum-weights}
\end{figure*}

The absolute branch probabilities in
Fig.~\ref{fig:qis-vacuum-weights} complement the normalized
branch entropies of Fig.~\ref{fig:vac-branch-qis}.
The exchange becomes sharper with mass and $\theta=pi$ corresponds to
rqual probabilities. Their variation with angle resolves how sharply the vacuum
redistributes flux as the mass increases.

\begin{figure}[t]
 \centering
 \includegraphics[width=\columnwidth]{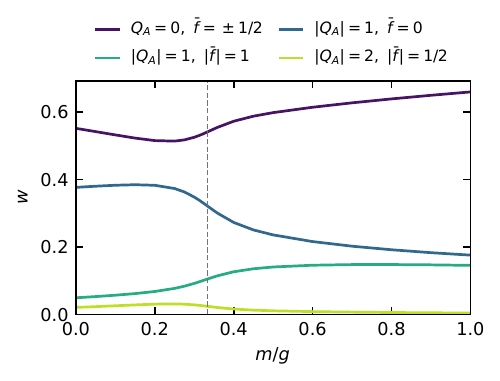}
 \caption{Vacuum sector probabilities at $\theta=\pi$
 for a half-space interval, $N=160$ and $ga=1/8$.
 Each curve sums sectors with the indicated
 $(|Q_A|,|\bar f|)$. The dashed line gives
 $m/g=0.3335$ as a reference near the critical mass.}
 \label{fig:qis-vacuum-sectorweights}
\end{figure}

The other flux sectors can be organized by the enclosed charge and the mean boundary field.
At $\theta=\pi$, the main sectors are $f_{L,R}=\ell_{L,R}+1/2$.
Then $Q_A=f_R-f_L$ and $\bar f=(f_L+f_R)/2$.
Figure~\ref{fig:qis-vacuum-sectorweights} groups their
weights by $(|Q_A|,|\bar f|)$.
Charged sectors retain appreciable probability across
the mass scan. Conditioning on the two neutral branches
therefore selects only part of the spatial state,
even where their exchange is sharp.

\begin{figure}[t]
 \centering
 \includegraphics[width=\columnwidth]{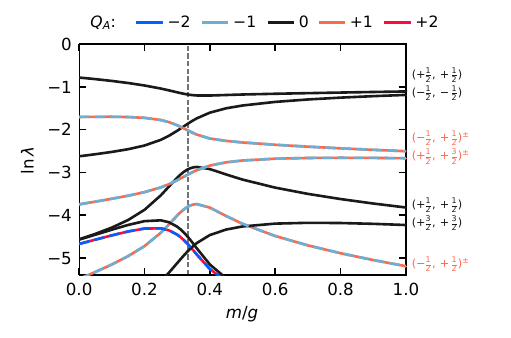}
 \caption{Leading entanglement spectrum at $\theta=\pi$,
 with $N=160$, $ga=1/8$ and a half-space interval.
 Color indicates $Q_A$, with solid lines for $Q_A\geq0$
 and dashed lines for $Q_A<0$.
 Labels at the right give $(f_L,f_R)$, with $\pm$
 denoting the two orderings of a reflected pair.
 The dashed vertical line gives $m/g=0.3335$.
}
 \label{fig:qis-vacuum-spectrum}
\end{figure}

Figure~\ref{fig:qis-vacuum-spectrum} resolves the dominant
eigenvalues of the same half-ring density matrix.
Here the plotted $\lambda$ denotes a full eigenvalue
$w_k\lambda_{ki}$, including its sector probability.
Reflection about the cut interchanges $(f_L,f_R)$ and
$(f_R,f_L)$ and reverses the enclosed charge, producing
the nearly coincident opposite-charge trajectories.
The neutral levels also rearrange near the critical
mass. This level redistribution complements the entropy
and flux-resolved nonstabilizerness in
Fig.~\ref{fig:qis-massscan}.

\subsection{Flux-resolved Bell fidelity}
\label{app:qis-bell}

The shared sector bases in Eq.~\eqref{eq:qis-block-assembly} make
the two-sided records explicit. After selecting $I,B$, a Schmidt
decomposition in each sector gives normalized probabilities
$\lambda_{Ii}$ and $\lambda_{Bi}$. Local isometries conditioned on
the flux record align the Schmidt bases with common residual bases
$|i\rangle_A,|i\rangle_{\bar A}$, giving
\begin{align}
 |\Psi_{IB}\rangle\longmapsto{}&
 \sqrt{p}\,|00\rangle\sum_i\sqrt{\lambda_{Ii}}\,|ii\rangle
 \nonumber\\
 &+e^{i\phi}\sqrt{1-p}\,|11\rangle
             \sum_i\sqrt{\lambda_{Bi}}\,|ii\rangle .
 \label{eq:qis-bell-aligned}
\end{align}
The first register on each side stores flux; the remaining registers
contain the within-sector states. Ranks may differ, in which case
the shorter spectrum is padded with zeros. This construction uses
only sector projectors and the pure-state Schmidt decomposition,
without assigning a complete intact or broken configuration.

For a normalized residual reference $|\chi\rangle=\sum_i c_i|ii\rangle$,
the overlap with Eq.~\eqref{eq:qis-bell-reference} is
\begin{equation}
 \sum_i c_i^* v_i,\qquad
 v_i=\frac{\sqrt{p\lambda_{Ii}}
               +\sqrt{(1-p)\lambda_{Bi}}}{\sqrt2}.
 \label{eq:qis-bell-overlap}
\end{equation}
Aligning the ordered Schmidt bases maximizes each branch overlap.
Cauchy--Schwarz then fixes the optimal residual coefficients to
$c_i=v_i/(\sum_j|v_j|^2)^{1/2}$. The optimum squared overlap is
$\sum_i|v_i|^2$, giving
Eq.~\eqref{eq:qis-bell-fidelity}. The conditions for unit fidelity
are $p=1/2$ and $\lambda_{Ii}=\lambda_{Bi}$ for every $i$.
Equal entropy alone is insufficient. The entropy identity
Eq.~\eqref{eq:qis-branch-maximal-entropy}, in contrast, follows
directly from biorthogonality~\cite{Linden:2005xuc} and
does not require matching internal spectra.

A leading spectral prefix suffices for bounds. If the first $r$
probabilities are known, define
\begin{equation}
 \beta_r=\sum_{i=1}^r\sqrt{\lambda_{Ii}\lambda_{Bi}},\qquad
 t_k=1-\sum_{i=1}^r\lambda_{ki}.
 \label{eq:qis-bell-prefix}
\end{equation}
Positivity and Cauchy--Schwarz on the remaining tails give
\begin{equation}
 \beta_r\leq\beta\leq\beta_r+\sqrt{t_I t_B}.
 \label{eq:qis-bell-tail-bound}
\end{equation}
Flux-resolved entropies also give bounds when individual eigenvalues
are unavailable. If $S_k<\ln2$, the largest eigenvalue exceeds
one half and satisfies $S_k\geq h_2(\lambda_{k1})$. $a_k$
is the root of $h_2(a_k)=S_k$ in $[1/2,1]$. Monotonicity on
this interval then gives $\lambda_{k1}\geq a_k$ and
\begin{equation}
 \sqrt{a_Ia_B}\leq\beta\leq1.
 \label{eq:qis-bell-entropy-bound}
\end{equation}
Together with Eq.~\eqref{eq:qis-bell-fidelity}, this bounds the
Bell fidelity without assigning unmeasured eigenvalues.
The lower bound is
$F_{\rm Bell}^{\rm lb}=1/2+\sqrt{p(1-p)a_Ia_B}$.
For a one-site interval, Gauss's law fixes the occupation within
each boundary-flux sector, so $S_I=S_B=0$ and $\beta=1$.
Figure~\ref{fig:qis-bell-fidelity} uses the exact fidelity in this
case. For larger intervals, it combines the entropy and spectral-prefix
bounds wherever both are available, taking the larger lower bound
and the smaller upper bound. Each bound is evaluated before averaging
over the two boundary alignments.

\section{Two-region density matrices and numerical conventions}
\label{app:spatial}

\subsection{Spin and fermionic reduced density matrices}
The interval decomposition is first $A$, then the gap $G$,
and $B$. The spin partial trace retains matrix
elements at fixed outer bounding fluxes and fixed gap occupation.
The charges of $A$ and $B$ may differ between the bra and ket
by opposite amounts. Gauss's law then changes the electric field
between them while preserving the outer fields. Fixing all four
bounding fluxes instead would remove these elements.

The matrices $\Gamma_{\rm e}$ and $\Gamma_{\rm o}$ are positive and not normalized, obtained by summing the retained blocks with even and odd physical occupation $N_G$ in the gap. All outer-flux labels are summed in each matrix. With $Z=\operatorname{tr}(\Gamma_{\rm e}+\Gamma_{\rm o})$, the two
reductions are
\begin{align}
 \rho_{AB}^{\rm D}
 &=\frac{\Gamma_{\rm e}+\Gamma_{\rm o}}{Z},
 \label{eq:spatial-rho-d}\\
 \rho_{AB}^{\rm F}
 &=\frac{\Gamma_{\rm e}+P_A\Gamma_{\rm o}P_A}{Z},
 \qquad P_A=(-1)^{N_A}.
 \label{eq:spatial-rho-f}
\end{align}
The main text uses $\rho_{AB}=\rho_{AB}^{\rm D}$.
The superscript D denotes the gap-parity dressing in fermionic variables and F denotes the fermionic reduced density matrix (rdm). The blocks are not normalized separately. The conjugation in
Eq.~\eqref{eq:spatial-rho-f} supplies the exchange sign when
the retained fermion modes are placed next to each other in the
reduced occupation basis. Because this sign depends on the
discarded gap occupation, the two summed matrices need not be
related by a local basis change.

Both rdms have the same local states, charge probabilities
and separately charge-dephased matrix:
\begin{equation}
 \rho_A^{\rm D}=\rho_A^{\rm F},\qquad
 \rho_B^{\rm D}=\rho_B^{\rm F},\qquad
 \mathcal D\rho_{AB}^{\rm D}=\mathcal D\rho_{AB}^{\rm F}.
 \label{eq:spatial-common-data}
\end{equation}
They differ only in entries with an odd change of regional occupation. For an empty gap they coincide. The equality of
the dephased matrices also makes $I_{\rm deph}$ identical; differences in mutual information arise entirely from $\DSasym$.

For a representative pair of sites $a\in A$ and $b\in B$ at the inner edges, define
\begin{equation}
 T_{ab}=\chi_a^\dagger W_{ab}\chi_b,\qquad
 P_G=(-1)^{N_G}.
 \label{eq:spatial-operator-dictionary}
\end{equation}
Here $W_{ab}$ is the Wilson line along the interval configuration. The corresponding coherence is $\langle T_{ab}\rangle$ for
the ordinary fermionic algebra and $\langle P_GT_{ab}\rangle$
for the spin algebra. Both are gauge invariant.
To see the distinction, write
$C_{\rm e,o}=\langle\Pi^G_{\rm e,o}T_{ab}\rangle$, where
$\Pi^G_{\rm e,o}$ project onto even and odd gap occupation.

The fermionic correlation is $C_{\rm e}+C_{\rm o}$ and the spin correlation is $C_{\rm e}-C_{\rm o}$. Because $C_{\rm e,o}$ are defined from the fermionic operator $T_{ab}$, $C_{\rm o}$ carries the exchange sign of the odd gap occupation. In the occupation basis the ground state is positive, so $C_{\rm e}$ and $-C_{\rm o}$ have the same sign. The spin correlation adds
the even and odd gap contributions and the fermionic correlation subtracts them, $|\langle T_{ab}\rangle|\le|\langle P_GT_{ab}\rangle|$.
The spin correlation is thus the charge transfer summed over all gap configurations and the fermionic correlation is weighted by the parity of the gap charge. The latter vanishes when the two parity sectors contribute equally, which requires equal probabilities of even and odd gap occupation and the same transfer amplitude in both sectors. Gauge invariance permits both choices.

For the orthogonal dephasing in
Eq.~\eqref{eq:spatial-dephasing}, the entropy increase equals a relative entropy~\cite{Gour:2009hng}.
The operator $\log(\mathcal D\rho)$ is block diagonal in the
dephasing sectors, so
$\operatorname{tr}[\rho\log(\mathcal D\rho)]
=\operatorname{tr}[\mathcal D\rho\log(\mathcal D\rho)]$.
Consequently,
\begin{align}
 D(\rho\Vert\mathcal D\rho)
 &=\operatorname{tr}(\rho\log\rho)
   -\operatorname{tr}[\rho\log(\mathcal D\rho)]
   \nonumber\\
 &=S(\mathcal D\rho)-S(\rho)\geq0.
 \label{eq:spatial-relative-entropy}
\end{align}
Because the marginal states commute with their local charges, dephasing leaves both marginal entropies unchanged and gives Eq.~\eqref{eq:spatial-mi-split}. Dephasing only $Q_A+Q_B$ would leave the entire joint state unchanged.

At separated probes, the vacuum-subtracted $I_{\rm deph}$ is small compared with the charge-coherence contribution for both $m/g=0.33$ and $0.50$. The dephased contribution matters more
for touching probes, as shown in Fig.~\ref{fig:spatial-halfcharge}.

\subsection{Comparison for the string}
\label{app:spatial-fermionic}

\begin{figure*}[tbp]
\centering
\includegraphics[width=\textwidth]{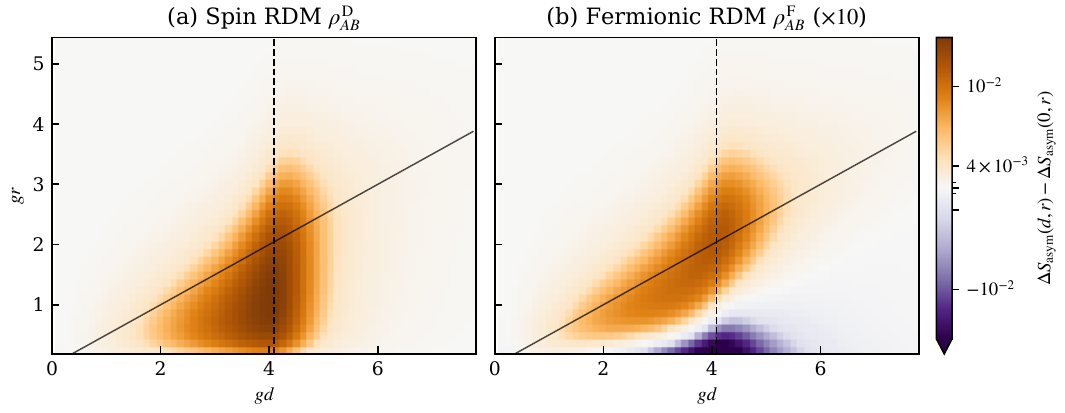}
\caption{Vacuum-subtracted charge coherence of the $q=1$ string at
$m/g=0.50$, $\theta=0$, $N=320$ and $ga=1/16$.
(a) Spin RDM $\rho_{AB}^{\rm D}$.
(b)  fermionic $\rho_{AB}^{\rm F}$, multiplied
by ten. The plotted quantity is
$\DSasym(d,r)-\DSasym(0,r)$.
Region $A$ lies at the midpoint and
$B$ is displaced by $r$; the solid line marks $r=d/2$.
The dashed vertical lines mark equal weights of the two dominant
central flux sectors.}
\label{fig:spatial-fermionic-ds}
\end{figure*}

Figure~\ref{fig:spatial-fermionic-ds} compares the two constructions at $m/g=0.50$. The response for the fermionic rdm is smaller, while the one for the spin rdm shows an extended enhancement near breaking. Since their charge probabilities and local states agree exactly, the difference
comes from how coherence is correlated with the charge in the gap.

\section{Neural-state construction and validation}
\label{app:neural-details}

\subsection{Parameter sharing and optimization}

The embedding in Eq.~\eqref{eq:neural-rbm} covers the integer range
$|L_n|\leq L_{\max}+N/2$, which contains every reconstructed integer flux allowed
by the neutral basis and reference cutoff. Its table has
$2L_{\max}+N+1$ rows and four complex entries per row.
For the vacuum, label hidden units by a filter $f=1,\ldots,8$
and its translation $t=0,\ldots,N/2-1$. The parameter constraints
are
\begin{equation}
 \begin{gathered}
 \boldsymbol W_{(f,t)n}=\boldsymbol W_{f,(n-2t)\bmod N},\\
 b^{\rm h}_{(f,t)}=b^{\rm h}_f,\qquad
 \boldsymbol b^{\rm v}_n=\boldsymbol b^{\rm v}_{n\bmod2}.
 \end{gathered}
 \label{eq:neural-sharing}
\end{equation}
A two-site translation permutes the hidden factors and leaves
their product unchanged. This invariance relates configurations whose
translations remain within the retained reference register. The string
network treats the left-hand parameters as independent. Assigning them the values on the
right gives the exact vacuum initialization used in
Sec.~\ref{sec:string-rbm}.

The implementation uses NetKet's physical basis and Hamiltonian construction with a custom complex RBM~\cite{Vicentini:2021pcv}. Before exponentiation we subtract the maximum real log-amplitude, which rescales the entire vector without changing the Rayleigh quotient. The vacuum calculation translates the small coupling tensor instead of replicating the full array of embedded configurations. The string calculation evaluates amplitudes in chunks. Both use exact sums and automatic differentiation through the sparse Hamiltonian.

For the $N=24$, $ga=1$ vacuum scans, each mass starts at $\theta=0$ from the optimized $N=24$, $ga=1/2$, $m/g=0.33$
vacuum used to initialize the string calculation. This first stage uses $1500$ Adam steps at learning rate $5\times10^{-3}$. The remaining angles use $800$ steps, initialized from the nearest completed angle at the same mass. Each state is then
refined in two rounds of $500$ Adam steps at $3\times10^{-4}$, with fresh optimizer variables. The second round starts from the lowest-energy parameters recorded in the first. ED overlaps are used only for validation.

\bibliography{refs}
\end{document}